\documentclass[aip,jcp,reprint,amsmath,amssymb,superscriptaddress,longbibliography,floatfix]{revtex4-2}

\usepackage{color}

\usepackage{bm}
\usepackage{physics}
\usepackage{placeins}
\usepackage{mathtools}
\usepackage{etoolbox}

\makeatletter
\def\@email#1#2{%
 \endgroup
 \patchcmd{\titleblock@produce}
  {\frontmatter@RRAPformat}
  {\frontmatter@RRAPformat{\produce@RRAP{*#1\href{mailto:#2}{#2}}}\frontmatter@RRAPformat}
  {}{}
}%
\makeatother

\begin{document}

\title{\textcolor{black}{Formulation of Brownian chain molecular dynamics and the path-resolved quasi-density operator}}

\author{Motoyuki Shiga}

\email{shiga.motoyuki@jaea.go.jp}

\affiliation{Center for Computational Science and e-Systems, Japan Atomic Energy Agency (JAEA), 178-4-4, Wakashiba, Kashiwa, Chiba, 277-0871, Japan}

\date{\today}

\begin{abstract}
Brownian chain molecular dynamics (BCMD) [M. Shiga, {\it J. Comput. Chem.} {\bf 43}, 1864 (2022)]
was introduced as a practical path-integral-based approach to approximate quantum
vibrational dynamics in molecular and condensed-phase systems.
Its equations of motion combine Newtonian dynamics of the centroid variables with
overdamped Brownian dynamics of the non-centroid path variables.
\begingroup\color{black}
Here, we introduce a path-resolved quasi-density-operator (path-QDO)
representation by extending the QDO construction underlying centroid molecular
dynamics to the full discretized imaginary-time path.
For two position-diagonal observables, this representation gives an exact expression
for the Kubo-transformed correlation function in the fine-discretization limit, without
requiring either observable to be linear.
BCMD can be viewed as a phenomenological, time-homogeneous Markov approximation
to the exact reduced propagation in these path variables, although its generator
is not derived from the path-QDO formulation.
\endgroup
We also formulate \textcolor{black}{and validate} a multiple-time-step algorithm that allows pseudo-Hamiltonian errors
in the deterministic propagation to be controlled by separating the Brownian refresh
interval from the inner integration time step.
\begingroup\color{black}
We test the path-QDO representation on one-dimensional models and present BCMD
infrared spectra of q-TIP4P/F ice, ambient water, and hot water, comparing them
with published results from other approximate quantum-dynamics methods.
\endgroup
\end{abstract}


\maketitle

\begin{center}
\footnotesize
This manuscript has been accepted for publication in
\textit{The Journal of Chemical Physics}.
\end{center}

\section{Introduction}

Imaginary-time path integrals map the quantum Boltzmann distribution onto an
extended classical system, usually represented as a ring polymer, and provide
the basis for path-integral Monte Carlo and path-integral molecular dynamics
(PIMD). They are now standard tools for including nuclear quantum effects in
equilibrium simulations of molecular and condensed-phase systems.
\cite{FeynmanHibbs1965,ChandlerWolynes1981,ParrinelloRahman1984,
Ceperley1995,MarxParrinello1996,MarxHutter2009,
MarklandCeriotti2018,Tuckerman2023,CeriottiManolopoulosMarklandRossi2026}

Several trajectory approximations use the same imaginary-time information to
approximate Kubo-transformed real-time correlation functions. Centroid
molecular dynamics (CMD) propagates the physical centroid on a potential of
mean force generated by the remaining path variables.
\cite{CaoVoth1994II,CaoVoth1994III,JangVoth1999CMD}
Ring-polymer molecular dynamics (RPMD) instead propagates the full ring polymer
with inertial dynamics in the extended space.
\cite{CraigManolopoulos2004,BraamsManolopoulos2006,Habershon2013RPMDReview,CeriottiManolopoulosMarklandRossi2026}
Partially adiabatic CMD (PACMD) is a centroid-dynamics formulation in which
the fictitious masses of the non-centroid modes are chosen to realize an
adiabatic separation, while thermostatting is used to sample the fast internal
modes.
Thermostatted RPMD (TRPMD) retains inertial ring-polymer internal modes and
couples them to a Langevin thermostat, with mode-dependent friction typically
chosen relative to the corresponding internal-mode frequencies and usually in
the underdamped regime.
\cite{RossiCeriottiManolopoulos2014}
In contrast, BCMD assigns no inertial dynamics to the non-centroid modes:
their first-order Brownian dynamics is defined by a common relaxation time
\(\tau_\alpha=\beta\hbar\), independent of the ring-polymer normal-mode
frequency.
Matsubara dynamics provides a formal framework for understanding the relations
among several of these approximations.
\cite{HeleWillattMuoloAlthorpe2015Matsubara,
HeleWillattMuoloAlthorpe2015Relation,Althorpe2021Review,Althorpe2024Review}

For vibrational spectroscopy, two well-known difficulties are especially
relevant here. Cartesian CMD can develop the curvature problem at low
temperature,
\cite{WittIvanovShigaForbertMarx2009,IvanovWittShigaMarx2010,
TreninsAlthorpe2018}
whereas RPMD can exhibit spurious resonances between physical vibrations and
the inertial internal modes of the ring polymer.
\cite{WittIvanovShigaForbertMarx2009,RossiCeriottiManolopoulos2014}
TRPMD suppresses these sharp resonances by thermostatting the internal modes,
although spectral broadening may result.
\cite{RossiCeriottiManolopoulos2014,RossiLiuPaesaniBowmanCeriotti2014}
A broadly applicable method that avoids these artifacts without introducing
system-specific dynamical parameters is therefore desirable.

More recent approaches include quasicentroid molecular dynamics (QCMD),
\cite{TreninsWillattAlthorpe2019,TreninsHaggardAlthorpe2022}
fast QCMD,
\cite{FletcherZhuLawrenceManolopoulos2021,
LawrenceLieberherrFletcherManolopoulos2023}
hybrid CMD and bead--Fourier CMD,
\cite{LimbuLondonFaruqueMomeni2025HCMD,LondonMomeni2025BFCMD}
temperature-elevated path-integral coarse-graining simulations (Te-PIGS),
\cite{MusilZaporozhetsNoeClementiKapil2022}
partially adiabatic elevated-temperature centroid molecular dynamics
(PA-\(T_e\)-CMD),\cite{CastroTreninsKapilRossi2025}
and constant-uncertainty molecular dynamics (CUMD).
\cite{Hasegawa2023CUMD}
Several of these methods are designed, in different ways, to mitigate the
curvature problem of CMD, which arises when the centroid free-energy
description is distorted by extended imaginary-time configurations,
particularly at low temperature. For molecular vibrational spectroscopy,
rotationally extended path configurations can make this effect especially
severe. QCMD addresses the problem by introducing curvilinear quasicentroid
coordinates, although their construction is system dependent and can become
nontrivial for large-amplitude motion. Fast QCMD retains the same basic
strategy but replaces the on-the-fly quasicentroid potential of mean force by
a fitted effective correction.
\cite{TreninsWillattAlthorpe2019,TreninsHaggardAlthorpe2022,
FletcherZhuLawrenceManolopoulos2021,
LawrenceLieberherrFletcherManolopoulos2023}
Te-PIGS takes a different route by constructing the quantum free-energy
correction at an elevated temperature, where the curvature effect is reduced,
and transferring this correction to lower temperatures.
\cite{MusilZaporozhetsNoeClementiKapil2022,
LawrenceLieberherrFletcherManolopoulos2023}

Other trajectory-based routes include equilibrium Liouville dynamics,
\cite{LiuMiller2011ELD,LiuMiller2011ELDII}
and path-integral Liouville dynamics.
\cite{Liu2014PILD,LiuZhang2016PILD}
Open-chain imaginary-time path-integral sampling provides another route to
approximate symmetrized quantum time-correlation functions.
\cite{CendagortaBacicTuckerman2018}
Influence-functional approaches provide numerically exact reduced quantum
dynamics for suitable system--bath formulations,
\cite{MakriMakarov1995,Makri1999Review}
while forward--backward semiclassical and quantum-trajectory constructions
offer related real-time approaches.
\cite{MakriNakayamaWright2004,NakayamaMakri2005,Makri2011Review}
Semiclassical initial-value-representation methods constitute another broad
class of trajectory-based approximations.
\cite{HermanKluk1984,KaledinMiller2003,LiuMiller2007LSC,
Liu2015LSCReview,HabershonManolopoulos2009ZPE,
BuchholzEtAl2018HK,ConteAietaCeotto2026Review}
Fully quantum wavefunction methods such as multiconfiguration time-dependent
Hartree (MCTDH) and its matrix-product-state formulation (MPS-MCTDH) provide
important benchmarks, although their routine application to large molecular
systems and condensed phases remains demanding.
\cite{BeckMeyerWorthMeyer2000MCTDH,Kurashige2018MPSMCTDH}

Brownian chain molecular dynamics (BCMD) was introduced as a distinct
approximation that eliminates the RPMD chain-resonance mechanism while
reducing the severity of the CMD curvature problem.
\cite{Shiga2022BCMD}
The centroid follows Newtonian dynamics, whereas the non-centroid path
variables undergo first-order overdamped Brownian dynamics. Since these
variables have no inertial normal-mode frequencies, the RPMD internal-mode
resonance mechanism is absent by construction. This distinguishes BCMD from
PACMD, in which the non-centroid modes retain auxiliary inertial dynamics with
fictitious masses chosen to realize an adiabatic separation, while
thermostatting is used to sample the fast internal modes.
\cite{HoneRosskyVoth2006,HabershonFanourgakisManolopoulos2008}
Both the mass scaling and the thermostatting protocol are therefore part of
the numerical specification of a PACMD calculation.
BCMD, by contrast, propagates the full set of discretized-path coordinates,
with Newtonian dynamics for the centroid and first-order Brownian dynamics for
the non-centroid coordinates. It is likewise distinct from TRPMD, which
retains inertial ring-polymer modes and couples the non-centroid modes to a
Langevin thermostat.
\cite{RossiCeriottiManolopoulos2014}

BCMD has so far been tested for several molecular and condensed-phase
systems. The original BCMD study examined simple model systems, gas-phase
molecules, protonated water clusters, and liquid water, and compared the
resulting spectra with exact results where available and with established
trajectory approximations.
\cite{Shiga2022BCMD}
For the systems examined there, BCMD reproduced the principal vibrational
features while avoiding the RPMD chain-resonance structure and alleviating
the curvature-related distortions found in standard CMD. The method was
subsequently applied to hydrogen in palladium using a neural-network potential
of density-functional-theory accuracy.
\cite{ShigaThomsenKimizuka2024PRB}
The calculated inelastic-neutron-scattering spectrum reproduced both the
fundamental excitation and the first overtone, including their positions and
relative intensities, substantially better than classical molecular dynamics
(MD) and in close agreement with experiment. This application demonstrates
the performance of BCMD for a strongly anharmonic quantum impurity in a
crystalline solid, beyond the molecular and liquid-water benchmarks of the
original work. BCMD has been implemented in the open-source PIMD
package\cite{PIMDweb} and, more recently, has also been incorporated as a
path-integral propagator in CP2K.\cite{CP2K2026BCMD}

The original proposal of BCMD was primarily algorithmic, whereas CMD has a
formal foundation in the quasi-density-operator (QDO) formulation of Jang and
Voth.
\cite{JangVoth1999QDO,JangVoth1999CMD}
The centroid-QDO formulation establishes a connection between centroid
dynamics and exact Kubo-transformed correlation functions. In its standard
form, however, this direct relation applies when at least one of the two
operators is linear in position and/or momentum.
\cite{GevaShiVoth2001}
Correlations involving general nonlinear operators can instead be treated
through higher-order constructions or generalized centroid dynamics (GCD),
in which additional centroids associated with selected composite operators
are introduced.
\cite{ReichmanRoyJangVoth2000,KrishnaVoth2006,Horikoshi2012GCD}

Here, we extend the QDO construction from centroid variables to the full
discretized imaginary-time path. We introduce a path-resolved quasi-density
operator (path-QDO) that resolves the imaginary-time position slice by slice
while retaining the centroid momentum. For two position-diagonal observables,
the resulting formulation yields an exact representation of the
Kubo-transformed correlation function in the fine-discretization limit without
requiring either observable to be linear. \begingroup\color{black}
Because BCMD propagates the full
discretized path rather than the centroid alone, the path-QDO provides an exact
reference representation in corresponding path variables. BCMD can be viewed as a phenomenological, time-homogeneous Markov
approximation to the formally exact reduced propagation.
\endgroup

The present formulation involves two conceptually distinct
steps. First, the path-QDO provides the exact path-resolved reference
\begingroup\color{black}
representation of the Kubo-transformed dynamics. Second, BCMD is introduced as an
approximate dynamical closure in the corresponding path variables.
\endgroup The path-QDO construction
does not uniquely determine the BCMD generator. The primary aim of the present
work is to clarify the theoretical framework in which BCMD is formulated. The
present work also examines its numerical performance relative to other
path-integral dynamical methods.

We next address a practical issue concerning the numerical stability of
long-time BCMD simulations. In the velocity-refresh representation used for
numerical integration, the BCMD algorithm naturally involves two distinct time
scales: an inner deterministic integration step, denoted by $\delta t$, and a
Brownian refresh interval, $\Delta t$, at which the auxiliary velocities
associated with the non-centroid coordinates are randomized. In the original
implementation of BCMD, these two time scales were taken to be identical,
$\delta t=\Delta t$, so that each deterministic propagation step was
immediately followed by a Brownian refresh.
\cite{Shiga2022BCMD}
While this simple choice is sufficient for relatively short trajectories, it
can become restrictive in long simulations because the time step required for
stable and accurate deterministic integration need not coincide with the
Brownian relaxation interval required by the BCMD dynamics.

Here we formulate a multiple-time-step (MTS) implementation of BCMD in which
the Brownian refresh interval $\Delta t$ is divided into several shorter
deterministic steps $\delta t$, with $\Delta t=L\delta t$. The deterministic
propagation over each interval $\Delta t$ is constructed using a symmetric
reversible splitting based on the reversible reference system propagator
algorithm (RESPA).
\cite{TuckermanBerneMartyna1992}
This separation allows $\delta t$ to be reduced to control the numerical error
of the deterministic propagation without changing the intended Brownian
refresh interval $\Delta t$. Because the stochastic refresh deliberately
changes the auxiliary kinetic energy, numerical stability is assessed from
the conservation of the pseudo-Hamiltonian during the deterministic
propagation between successive refresh events. We show below that this MTS
formulation substantially improves the stability of long-time BCMD
simulations by allowing the deterministic integration accuracy to be
controlled independently of the Brownian relaxation time scale.

Finally, we extend the numerical tests beyond those of the original BCMD
study. Using the q-TIP4P/F water model,
\cite{HabershonMarklandManolopoulos2009}
we calculate infrared spectra for ice at 150 K, ambient liquid water at
300 K, and high-temperature dense water at 600 K. These state points have
previously been studied with several other approximate quantum-dynamics
methods,
\cite{RossiLiuPaesaniBowmanCeriotti2014,TreninsWillattAlthorpe2019,
BensonTreninsAlthorpe2020}
so the new element here is the assessment of BCMD itself across phase and
temperature.

The remainder of this paper is organized as follows. Section~II develops the
path-QDO formulation and introduces the BCMD generator. Section~III discusses
the basic properties of BCMD. Section~IV presents the velocity-refresh
representation and the MTS algorithm. Section~V reports the numerical tests
of the path-QDO representation, the MTS integrator, and the q-TIP4P/F
spectra. Section~VI summarizes the conclusions.


\begingroup\color{black}
\section{Path-QDO representation and BCMD approximation}
\endgroup

\subsection{Path-resolved quasi-density operator}

We consider a one-dimensional quantum system with Hamiltonian
\begin{equation}
\hat H = \frac{\hat p^{2}}{2m} + V(\hat x),
\label{eq:H_1d}
\end{equation}
and canonical partition function
\begin{equation}
Z = {\rm Tr}\left(e^{-\beta\hat H}\right).
\label{eq:Z_exact}
\end{equation}
The Kubo-transformed correlation function of two operators $\hat A$ and
$\hat B$ is
\begin{equation}
C_{AB}^{\rm K}(t)
=
\frac{1}{\beta Z}
\int_0^\beta d\lambda\,
{\rm Tr}
\left[
e^{-(\beta-\lambda)\hat H}
\hat A
e^{-\lambda\hat H}
\hat B(t)
\right],
\label{eq:Kubo_correlation}
\end{equation}
where
$\hat B(t)=e^{i\hat Ht/\hbar}\hat B e^{-i\hat Ht/\hbar}$.

The quasi-density-operator (QDO) formulation provides the formal basis of
centroid molecular dynamics (CMD).
\cite{JangVoth1999QDO,JangVoth1999CMD}
In terms of the centroid phase-space variables $(q_{\rm c},p_{\rm c})$, the
non-normalized centroid QDO is defined as
\begin{align}
\hat\varphi_{\beta}^{\,{\rm CMD}}(q_{\rm c},p_{\rm c})
&=
\int\frac{d\eta}{2\pi}
\int\frac{d\nu}{2\pi\hbar}\,
e^{-i\eta q_{\rm c}}
e^{-i\nu p_{\rm c}/\hbar}
\nonumber\\
&\quad\times
e^{-\beta\hat H+i\eta\hat x+i\nu\hat p/\hbar}.
\label{eq:CMD_varphi_definition}
\end{align}
When $\hat A$ or $\hat B$
 is linear in position and/or momentum,
 Eq.~(\ref{eq:Kubo_correlation}) can  be reexpressed as a centroid phase-space
average.
\cite{JangVoth1999QDO,JangVoth1999CMD,GevaShiVoth2001}
CMD is obtained by approximating the time evolution of $\hat B(t)$ by classical propagation
of the centroid on its potential of mean force.
However, if neither operator is linear, the  Kubo correlation does not, in
general, close within the standard centroid variables, and generalized
centroid or higher-order constructions are required.
\cite{ReichmanRoyJangVoth2000}

In this paper, we extend the QDO construction from the centroid to the full discretized
imaginary-time path. For a Trotter number $P$, we introduce
\begin{equation}
\mathbf{x}\equiv(x_1,x_2,\ldots,x_P)^{\mathsf T},
\qquad
x_0\equiv x_P,
\qquad
x_{P+1}\equiv x_1,
\label{eq:path_notation}
\end{equation}
with cyclic bead indices.

The purpose of the path-QDO construction is to remove this operator-linearity
restriction for position observables.  Instead of projecting the
imaginary-time path onto \(q_{\rm c}\), we retain every slice coordinate
\(\mathbf{x}\).  We therefore introduce the path-QDO
\begin{align}
&\hat\varphi_{\beta,P}^{}(\mathbf{x},p_{\rm c}^{})
\equiv
\int\frac{d\nu^{}}{2\pi\hbar}
\prod_{s=1}^{P}\frac{d\eta_s^{}}{2\pi}
\prod_{s=1}^{P}
\Biggl[
 e^{i\eta_s^{}(\hat x^{}-x_s^{})}
\nonumber\\
&\quad\quad\quad\quad\times
 e^{i\nu^{}(\hat p^{}-p_{\rm c}^{})/(P\hbar)}
 e^{-\beta^{}\hat H^{}/P}
\Biggr].
\label{eq:path_QDO_definition}
\end{align}
The variables \(\eta_s\) constrain the position slice by slice.  Since
\(\prod_s e^{-i\nu p_{\rm c}/(P\hbar)}=e^{-i\nu p_{\rm c}/\hbar}\), the single
variable \(p_{\rm c}\) is conjugate to the centroid of the corresponding path
momenta.

The density associated with the path-QDO and its normalized form are defined as
\begin{equation}
\rho_{\beta,P}^{}(\mathbf{x},p_{\rm c}^{})
\equiv
{\rm Tr}\left[\hat\varphi_{\beta,P}^{}(\mathbf{x},p_{\rm c}^{})\right]
\label{eq:path_QDO_density}
\end{equation}
and
\begin{equation}
\hat\delta_{\beta,P}^{}(\mathbf{x},p_{\rm c}^{})
\equiv
\frac{\hat\varphi_{\beta,P}^{}(\mathbf{x},p_{\rm c}^{})}
{\rho_{\beta,P}^{}(\mathbf{x},p_{\rm c}^{})},
\label{eq:path_QDO_delta}
\end{equation}
respectively.
For any operator \(\hat A\), we define
\begin{equation}
A_{\beta,P}^{}(\mathbf{x},p_{\rm c}^{})
\equiv
{\rm Tr}\left[
\hat\delta_{\beta,P}^{}(\mathbf{x},p_{\rm c}^{})\hat A^{}
\right]
\label{eq:path_QDO_Aeff}
\end{equation}
and
\begin{equation}
A_{\beta,P}^{}(\mathbf{x},p_{\rm c}^{},t)
\equiv
{\rm Tr}\left[
\hat\delta_{\beta,P}^{}(\mathbf{x},p_{\rm c}^{})\hat A^{}(t)
\right].
\label{eq:path_QDO_exact_dynamics}
\end{equation}

The path-QDO  resolves the Boltzmann operator exactly at every
finite \(P\).  Indeed,
\begin{align}
\int dx_s^{}\,e^{-i\eta_s^{}x_s^{}}
&=2\pi\delta(\eta_s^{}),
\nonumber\\
\int dp_{\rm c}^{}\,e^{-i\nu^{}p_{\rm c}^{}/\hbar}
&=2\pi\hbar\delta(\nu^{}),
\label{eq:path_QDO_fourier_collapse}
\end{align}
so that
\begin{equation}
\int dp_{\rm c}^{}\int d\mathbf{x}\,
\hat\varphi_{\beta,P}^{}(\mathbf{x},p_{\rm c}^{})
=
\prod_{s=1}^{P}e^{-\beta^{}\hat H^{}/P}
=e^{-\beta^{}\hat H^{}}
\label{eq:path_QDO_integral_identity}
\end{equation}
where
\begin{equation}
d\mathbf{x}\equiv\prod_{s=1}^{P}dx_s^{}.
\label{eq:path_measure_definition}
\end{equation}
Taking the trace gives
\begin{equation}
Z^{}=
\int dp_{\rm c}^{}\int d\mathbf{x}\,
\rho_{\beta,P}^{}(\mathbf{x},p_{\rm c}^{}),
\label{eq:path_QDO_partition_identity}
\end{equation}
and hence
\begin{equation}
\langle\hat A^{}\rangle
=
\frac{1}{Z^{}}
\int dp_{\rm c}^{}\int d\mathbf{x}\,
\rho_{\beta,P}^{}(\mathbf{x},p_{\rm c}^{})
A_{\beta,P}^{}(\mathbf{x},p_{\rm c}^{}).
\label{eq:path_QDO_static_average}
\end{equation}
The exact quantum time evolution does not change this canonical statistical
average.  Indeed, using Eq.~\eqref{eq:path_QDO_integral_identity}, one obtains,
for every finite \(P\),
\begin{align}
&\frac{1}{Z^{}}
\int dp_{\rm c}^{}\int d\mathbf{x}\,
\rho_{\beta,P}^{}(\mathbf{x},p_{\rm c}^{})
A_{\beta,P}^{}(\mathbf{x},p_{\rm c}^{},t)
\nonumber\\
&\quad=
\frac{1}{Z^{}}
{\rm Tr}\left[e^{-\beta^{}\hat H^{}}\hat A^{}(t)\right]
=
\frac{1}{Z^{}}
{\rm Tr}\left[e^{-\beta^{}\hat H^{}}\hat A^{}\right]
=
\langle\hat A^{}\rangle .
\label{eq:path_QDO_static_average_preserved}
\end{align}
The middle equality follows from cyclic invariance of the trace and from the
commutation of \(e^{-\beta\hat H}\) with the real-time evolution operator.
Thus exact propagation changes the path-resolved estimator, but not its
canonical average.  This statement applies to the formal path-QDO and to an
arbitrary operator \(\hat A\).  

\subsection{Ring-polymer representation}

We now introduce the ring-polymer representation.
For each imaginary-time slice, the intermediate matrix element
following from Eq.~\eqref{eq:path_QDO_definition} is
\begin{align}
&\int\frac{d\eta_s}{2\pi}
\langle x_s'|
 e^{i\eta_s(\hat x-x_s)}
 e^{i\nu(\hat p-p_{\rm c})/(P\hbar)}
 e^{-\beta\hat H/P}
|x_{s+1}'\rangle
\nonumber\\
&\qquad=
\delta(x_s'-x_s)\,e^{-i\nu p_{\rm c}/(P\hbar)}
\langle x_s|
 e^{i\nu\hat p/(P\hbar)}e^{-\beta\hat H/P}
|x_{s+1}'\rangle
\nonumber\\
&\qquad=
\delta(x_s'-x_s)\,e^{-i\nu p_{\rm c}/(P\hbar)}
\langle x_s+\nu/P|e^{-\beta\hat H/P}|x_{s+1}'\rangle .
\label{eq:link_from_path_QDO}
\end{align}
Thus the factor \(e^{-i\nu p_{\rm c}/(P\hbar)}\) is extracted from each
link, and the product of the \(P\) factors gives the global Fourier factor
\(e^{-i\nu p_{\rm c}/\hbar}\) below.
For a closed path, the direct symmetric-Trotter representation is
\begin{equation}
\rho_{\beta,P}^{\,{\rm T}}(\mathbf{x},p_{\rm c}^{})
=
\int\frac{d\nu^{}}{2\pi\hbar}\,
e^{-i\nu^{}p_{\rm c}^{}/\hbar}
\prod_{s=1}^{P}g_P^{}(x_s^{},x_{s+1}^{};\nu^{})
\label{eq:rho_rp_link_product}
\end{equation}
using
\begin{align}
g_P^{}(a^{},b^{};\nu^{})
&\equiv
\left(\frac{m^{}P}{2\pi\beta^{}\hbar^{2}}\right)^{1/2}
\exp\left[
-\frac{m^{}P}{2\beta^{}\hbar^{2}}
\left(a^{}+\frac{\nu^{}}{P}-b^{}\right)^{2}
\right]
\nonumber\\
&\quad\times
\exp\left[
-\frac{\beta^{}}{2P}
\left\{V\!\left(a^{}+\frac{\nu^{}}{P}\right)+V(b^{})\right\}
\right].
\label{eq:rp_short_time_link}
\end{align}
The shifted potential in Eq.~\eqref{eq:rp_short_time_link}
prevents an exact separation of the centroid momentum at finite \(P\).  The
direct Trotter representation differs from the conventional ring-polymer
expression by
\begin{equation}
R_P(\mathbf{x},\nu)
\equiv
\exp\!\left[
-\frac{\beta}{2P}\sum_{s=1}^{P}
\left\{
V\!\left(x_s+\frac{\nu}{P}\right)-V(x_s)
\right\}
\right].
\label{eq:finite_P_potential_shift_factor}
\end{equation}
For fixed \(\nu\) and a smooth potential,
\begin{equation}
V\!\left(x_s+\frac{\nu}{P}\right)
=
V(x_s)+\frac{\nu}{P}V'(x_s)+O(P^{-2}),
\label{eq:finite_P_potential_shift_taylor}
\end{equation}
so that, provided the bead-averaged force remains finite,
\begin{equation}
\ln R_P
=
-\frac{\beta\nu}{2P^2}\sum_{s=1}^{P}V'(x_s)+O(P^{-2})
=O(P^{-1}).
\label{eq:finite_P_shift_vanishes}
\end{equation}
Moreover, cyclicity gives
\begin{equation}
\sum_{s=1}^{P}\left(x_s+\frac{\nu}{P}-x_{s+1}\right)^2
=
\sum_{s=1}^{P}(x_s-x_{s+1})^2+\frac{\nu^2}{P},
\label{eq:nu_shifted_spring_identity_corrected}
\end{equation}
so the kinetic part generates the \(P\)-independent Gaussian factor
\(\exp[-m\nu^2/(2\beta\hbar^2)]\).  Thus the characteristic scale of the
\(\nu\) integration remains finite as \(P\) increases, consistently with the
expansion in \(\nu/P\).  Under the usual integrability conditions of the
path-integral representation, \(R_P\to1\) as \(P\to\infty\).  I therefore use
below the conventional finite-\(P\) ring-polymer discretization, which has the
same continuum imaginary-time limit:
\begin{equation}
\rho_{\beta,P}^{\,{\rm rp}}(\mathbf{x},p_{\rm c}^{})
\equiv
M_{\beta}^{}(p_{\rm c}^{})\,
\bar\rho_{\beta,P}^{\,{\rm rp}}(\mathbf{x}),
\label{eq:rho_rp_pc_factorization_early}
\end{equation}
where
\begin{equation}
M_{\beta}^{}(p_{\rm c}^{})
\equiv
\left(\frac{\beta^{}}{2\pi m^{}}\right)^{1/2}
\exp\!\left[-\beta^{}\frac{p_{\rm c}^{2}}{2m^{}}\right]
\label{eq:Maxwell_pc_definition}
\end{equation}
is normalized to unity.
The Maxwell factor follows from the Gaussian Fourier
transform associated with Eq.~\eqref{eq:nu_shifted_spring_identity_corrected}
after the continuum-limit replacement described above.
The configurational density is
\begin{equation}
\bar\rho_{\beta,P}^{\,{\rm rp}}(\mathbf{x})
\equiv
\left(\frac{m^{}P}{2\pi\beta^{}\hbar^{2}}\right)^{P/2}
\exp\!\left[-\beta^{}V_{\rm eff}^{\,{\rm rp}}(\mathbf{x})\right],
\label{eq:ring_polymer_density_x}
\end{equation}
where
\begin{equation}
V_{\rm eff}^{\,{\rm rp}}(\mathbf{x})
=
\sum_{s=1}^{P}\frac{m^{}P}{2\beta^{2}\hbar^{2}}
(x_{s+1}^{}-x_s^{})^{2}
+\frac{1}{P}\sum_{s=1}^{P}V(x_s^{}).
\label{eq:Veff_beads_1d}
\end{equation}
Equivalently,
\begin{align}
V_{\rm eff}^{\,{\rm rp}}(\mathbf{x})
&=
\frac{1}{2}m^{}\omega_P^{2}\,\mathbf{x}^{\mathsf T}\mathbf{A}\mathbf{x}+\phi(\mathbf{x}),
\label{eq:Veff_matrix_1d}\\
\omega_P^{}
&\equiv
\frac{\sqrt{P}}{\beta^{}\hbar^{}},
\label{eq:omegaP_definition}\\
\phi(\mathbf{x})
&\equiv
\frac{1}{P}\sum_{s=1}^{P}V(x_s^{}),
\label{eq:phi_x_definition}
\end{align}
using the cyclic spring matrix
\begin{equation}
(\mathbf{A})_{ss'}
=
2\delta_{ss'}^{}-\delta_{s,s'+1}^{}-\delta_{s,s'-1}^{}.
\label{eq:A_matrix_definition}
\end{equation}
At finite \(P\),
Eqs.~\eqref{eq:rho_rp_pc_factorization_early}--\eqref{eq:Veff_beads_1d}
define the conventional ring-polymer discretization used below.  The direct
symmetric-Trotter evaluation of the path-QDO density contains the correction
factor in Eq.~\eqref{eq:finite_P_potential_shift_factor}; the difference
vanishes in the continuum imaginary-time limit.
The associated partition function
\begin{equation}
Z_P^{}
\equiv
\int d\mathbf{x}\,\bar\rho_{\beta,P}^{\,{\rm rp}}(\mathbf{x})
=
\int d\mathbf{x}\int dp_{\rm c}^{}\,
\rho_{\beta,P}^{\,{\rm rp}}(\mathbf{x},p_{\rm c}^{})
\label{eq:ZP_rp_definition_path_QDO}
\end{equation}
satisfies \(Z_P\to Z\) as \(P\to\infty\).\cite{CaoVoth1994I,CaoVoth1994III}

\subsection{Kubo-transformed correlations for position operators}

In the remainder of the formulation, both observables are restricted to
position-diagonal operators,
\begin{equation}
\hat A^{}=A(\hat x^{}),
\quad
\hat B^{}=B(\hat x^{}),
\label{eq:position_diagonal_AB_path_QDO}
\end{equation}
where \(A(x)\) and \(B(x)\) may be arbitrary nonlinear functions.
Their Kubo-transformed correlation function is
\begin{equation}
C_{AB}^{\,{\rm K}}(t)
\equiv
\frac{1}{\beta^{}Z^{}}
\int_{0}^{\beta}d\lambda^{}\,
{\rm Tr}\left[
 e^{-(\beta^{}-\lambda^{})\hat H^{}}
 A(\hat x^{})
 e^{-\lambda^{}\hat H^{}}
 \hat B^{}(t)
\right],
\label{eq:Kubo_definition_path_QDO}
\end{equation}
where
\begin{equation}
\hat B^{}(t)
\equiv
e^{i\hat H^{}t/\hbar}B(\hat x^{})e^{-i\hat H^{}t/\hbar}.
\label{eq:B_Heisenberg_unperturbed_path_QDO}
\end{equation}

Now we introduce a perturbation in the Hamiltonian
with additional source term $\epsilon^{}A(\hat x^{})$ as
\begin{equation}
\hat H_{\epsilon}^{}
\equiv
\hat H^{}+\epsilon^{}A(\hat x^{}).
\label{eq:source_Hamiltonian_path_QDO}
\end{equation}
We take the imaginary-time Boltzmann factors to contain \(\hat H_\epsilon\).
The real-time operator \(\hat B(t)\) remains evolved by the unperturbed
Hamiltonian \(\hat H\) defined in Eq.~\eqref{eq:B_Heisenberg_unperturbed_path_QDO}.
Applying the Duhamel identity to Eq.~\eqref{eq:Kubo_definition_path_QDO} gives
\begin{equation}
C_{AB}^{\,{\rm K}}(t)
=-\frac{1}{\beta^{}Z^{}}
\left.
\frac{\partial}{\partial\epsilon^{}}
{\rm Tr}\left[e^{-\beta^{}\hat H_{\epsilon}^{}}\hat B^{}(t)\right]
\right|_{\epsilon=0}.
\label{eq:Kubo_from_source_derivative}
\end{equation}
The perturbed path-QDO is then obtained from
Eq.~\eqref{eq:path_QDO_definition} by replacing \(\hat H\) in every
imaginary-time factor by \(\hat H_{\epsilon}\) as
\begin{align}
&\hat\varphi_{\beta,P,\epsilon}^{}(\mathbf{x},p_{\rm c}^{})
\equiv
\int\frac{d\nu^{}}{2\pi\hbar}
\prod_{s=1}^{P}\frac{d\eta_s^{}}{2\pi}
\prod_{s=1}^{P}
\Biggl[
 e^{i\eta_s^{}(\hat x^{}-x_s^{})}
\nonumber\\
&\quad\quad\quad\quad\times
 e^{i\nu^{}(\hat p^{}-p_{\rm c}^{})/(P\hbar)}
 e^{-\beta^{}\hat H_{\epsilon}^{}/P}
\Biggr].
\label{eq:source_path_QDO_definition}
\end{align}
To make the estimator explicitly invariant under imaginary-time
translation, we define the cyclic-shift matrix \(\mathbf{C}_r\) by
\begin{equation}
\left(\mathbf{C}_r\mathbf{x}\right)_s
\equiv x_{s+r}^{},
\quad
r=0,\ldots,P-1,
\quad
x_{s+P}^{}\equiv x_s^{}.
\label{eq:cyclic_shift_definition}
\end{equation}
Thus \(\mathbf{C}_0=\mathbf{I}\), and \(\mathbf{C}_r\mathbf{x}\) represents the same
closed path with a different choice of imaginary-time origin.  We then define a dynamical object
\begin{equation}
\widetilde B_{\beta,P,\epsilon}^{}(\mathbf{x},p_{\rm c}^{},t)
\equiv
\frac{1}{P}\sum_{r=0}^{P-1}
{\rm Tr}\left[
\hat\varphi_{\beta,P,\epsilon}^{}(\mathbf{C}_r\mathbf{x},p_{\rm c}^{})
\hat B^{}(t)
\right].
\label{eq:Btilde_source_definition}
\end{equation}
The average over \(r\) removes the arbitrary choice of the imaginary-time
origin, or equivalently the bond at which the closed path is cut when
\(\hat B(t)\) is inserted.  The 
identity similar to
Eq.~\eqref{eq:path_QDO_integral_identity}
therefore gives, for every finite \(P\),
\begin{equation}
\int d\mathbf{x}\int dp_{\rm c}^{}\,
\widetilde B_{\beta,P,\epsilon}^{}(\mathbf{x},p_{\rm c}^{},t)
=
{\rm Tr}\left[e^{-\beta^{}\hat H_{\epsilon}^{}}\hat B^{}(t)\right],
\label{eq:source_generating_function_path_QDO_exact}
\end{equation}
since a cyclic shift is only a permutation of the
integration variables, each term has the same integral.  

Next, we denote the representations
\begin{align}
\rho_{\beta,P,\epsilon}^{\,{\rm rp}}(\mathbf{x},p_{\rm c}^{})
&\equiv
\left[
{\rm Tr}\,\hat\varphi_{\beta,P,\epsilon}^{}(\mathbf{x},p_{\rm c}^{})
\right]_{\rm rp},
\label{eq:rho_source_rp_definition}\\
\widetilde B_{\beta,P,\epsilon}^{\,{\rm rp}}(\mathbf{x},p_{\rm c}^{},t)
&\equiv
\left[
\widetilde B_{\beta,P,\epsilon}^{}(\mathbf{x},p_{\rm c}^{},t)
\right]_{\rm rp}.
\label{eq:Btilde_source_rp_definition}
\end{align}
These representations are symmetric with respect to the \(P\) imaginary-time slices.
Within the conventional finite-\(P\) ring-polymer discretization
defined in Sec.~II B, the closed-path density for the position-diagonal
perturbation in Eq.~\eqref{eq:source_Hamiltonian_path_QDO} acquires the usual
bead factor,
\begin{equation}
\rho_{\beta,P,\epsilon}^{\,{\rm rp}}(\mathbf{x},p_{\rm c}^{})
=
e^{-\beta^{}\epsilon^{}\bar A_P^{\,{\rm rp}}(\mathbf{x})}
\rho_{\beta,P}^{\,{\rm rp}}(\mathbf{x},p_{\rm c}^{}),
\label{eq:rho_source_factorization}
\end{equation}
where
\begin{equation}
\bar A_P^{\,{\rm rp}}(\mathbf{x})
\equiv
\frac{1}{P}\sum_{s=1}^{P}A(x_s^{}),
\label{eq:Abar_bead_average_rp}
\end{equation}
and
\(\rho_{\beta,P}^{\,{\rm rp}}\equiv
\rho_{\beta,P,0}^{\,{\rm rp}}\).

Note here that one imaginary-time bond
 is open in Eq.~\eqref{eq:Btilde_source_rp_definition}.
In the term for
which the bond between \(x_{s-1}\) and \(x_s\) is cut, let \(z\) denote the
open endpoint connected to \(x_s\) by the kernel of \(\hat B(t)\).  The
 short-time representation then contains the perturbation factor
\begin{equation}
e^{-\beta^{}\epsilon^{}\bar A_P^{\,{\rm rp}}(\mathbf{x})}
\exp\left[
-\frac{\beta^{}\epsilon^{}}{2P}
\left\{A(z^{})-A(x_s^{})\right\}
\right].
\label{eq:open_source_factor}
\end{equation}
The second factor differs from unity only through the single open endpoint.
Applying the source derivative to Eq.~(\ref{eq:open_source_factor}) gives
\begin{align}
&-\frac{1}{\beta}
\left.
\frac{\partial}{\partial\epsilon}
\left[
e^{-\beta\epsilon\bar A_P^{\,{\rm rp}}(\mathbf{x})}
\exp\left(
-\frac{\beta\epsilon}{2P}
\{A(z)-A(x_s)\}
\right)
\right]
\right|_{\epsilon=0}
\nonumber\\
&\qquad =
\bar A_P^{\,{\rm rp}}(\mathbf{x})
+\frac{1}{2P}\{A(z)-A(x_s)\}.
\label{eq:open_source_full_derivative}
\end{align}
The second term is the open-endpoint correction which carries factor \(P^{-1}\), and
vanishes in the continuum imaginary-time limit.  The standard bead estimator
\(\bar A_P^{\,{\rm rp}}\) is therefore obtained as \(P\to\infty\).

At \(\epsilon=0\), we write
\(\widetilde B_{\beta,P}^{\,{\rm rp}}\equiv
\widetilde B_{\beta,P,0}^{\,{\rm rp}}\) and define the normalized
path-resolved estimator by
\begin{equation}
\bar B_{\beta,P}^{\,{\rm rp}}(\mathbf{x},p_{\rm c}^{},t)
\equiv
\frac{\widetilde B_{\beta,P}^{\,{\rm rp}}(\mathbf{x},p_{\rm c}^{},t)}
{\rho_{\beta,P}^{\,{\rm rp}}(\mathbf{x},p_{\rm c}^{})}.
\label{eq:Bbar_rp_definition}
\end{equation}
Using Eqs.~\eqref{eq:Kubo_from_source_derivative} and
\eqref{eq:source_generating_function_path_QDO_exact}, applying the symmetric
ring-polymer representations of Eqs.~\eqref{eq:rho_source_rp_definition} and \eqref{eq:Btilde_source_rp_definition}, and then taking \(P\to\infty\) finally gives
\begin{align}
C_{AB}^{\,{\rm K}}(t)
&=
\lim_{P\to\infty}
\frac{1}{Z_P^{}}
\int d\mathbf{x}\int dp_{\rm c}^{}\,
\rho_{\beta,P}^{\,{\rm rp}}(\mathbf{x},p_{\rm c}^{})
\nonumber\\
&\quad\times
\bar A_P^{\,{\rm rp}}(\mathbf{x})
\bar B_{\beta,P}^{\,{\rm rp}}(\mathbf{x},p_{\rm c}^{},t).
\label{eq:Kubo_path_QDO_final_rp}
\end{align}

Equation~\eqref{eq:Kubo_path_QDO_final_rp} is the key path-QDO closure.  The
first operator enters through the full-path bead estimator
\(\bar A_P^{\,{\rm rp}}(\mathbf{x})\), while all exact
real-time quantum evolution of the second operator is contained in the
path-resolved quantity
\(\bar B_{\beta,P}^{\,{\rm rp}}(\mathbf{x},p_{\rm c};t)\).  Because the
individual imaginary-time coordinates are retained rather than compressed to
the centroid, this identity remains valid for nonlinear \(A(x)\) and
\(B(x)\).  Thus, in the limit \(P\to\infty\), the ordinary Kubo-transformed
correlation of two arbitrary position-diagonal operators closes exactly in the
path variables \((\mathbf{x},p_{\rm c})\); neither operator is required to be
linear.

At $t=0$, it can be shown from the definition of
Eq.~\eqref{eq:Bbar_rp_definition} that
\begin{equation}
\bar B_{\beta,P}^{\,{\rm rp}}(\mathbf{x},p_{\rm c}^{},0)
=
\frac{1}{P}\sum_{s=1}^{P}B(x_s^{})
\equiv
\bar B_P^{\,{\rm rp}}(\mathbf{x}).
\label{eq:Bbar_initial_bead_average_rp}
\end{equation}
The initial estimator is therefore independent of \(p_{\rm c}\).  For
\(t>0\), the path-resolved estimator generally depends on both the full path
\(\mathbf{x}\) and the centroid momentum \(p_{\rm c}\).

\subsection{Normal-mode representation of the ring polymer}

We adopt the normal-mode convention used in the 
formulations of the PIMD family.\cite{WittIvanovShigaForbertMarx2009}  Let \(\mathbf{U}\) be a real orthogonal matrix that
diagonalizes the scaled spring matrix \(P\mathbf{A}\):
\begin{equation}
P\,\mathbf{A}\mathbf{U}=\mathbf{U}\boldsymbol{\Lambda},
\quad
\mathbf{U}^{\mathsf T}\mathbf{U}=\mathbf{I},
\quad
\boldsymbol{\Lambda}={\rm diag}(\lambda_1^{},\ldots,\lambda_P^{}).
\label{eq:PA_diagonalization}
\end{equation}
The normal coordinates and their inverse transformation are defined by
\begin{equation}
\mathbf{q}
\equiv
\frac{1}{\sqrt{P}}\mathbf{U}^{\mathsf T}\mathbf{x},
\quad
\mathbf{x}=\sqrt{P}\,\mathbf{U}\mathbf{q}.
\label{eq:normal_mode_transform_1d}
\end{equation}
The scaled diagonalization gives
\begin{equation}
\mathbf{x}^{\mathsf T}\mathbf{A}\mathbf{x}=\mathbf{q}^{\mathsf T}\boldsymbol{\Lambda}\mathbf{q}.
\label{eq:spring_quadratic_normal_modes}
\end{equation}
The first column of \(\mathbf{U}\) is \(U_{s1}=P^{-1/2}\), so
\begin{equation}
q_1^{}\equiv q_{\rm c}^{}
=\frac{1}{P}\sum_{s=1}^{P}x_s^{}.
\label{eq:centroid_definition_1d}
\end{equation}
We write
\begin{equation}
\mathbf{q}\equiv(q_{\rm c}^{},q_2^{},\ldots,q_P^{})^{\mathsf T},
\label{eq:q_vector_definition_1d}
\end{equation}
and
\begin{equation}
x_s^{}(\mathbf{q})
=\sqrt{P}\sum_{\alpha=1}^{P}U_{s\alpha}^{}q_{\alpha}^{}
=q_{\rm c}^{}+\sqrt{P}\sum_{\alpha=2}^{P}U_{s\alpha}^{}q_{\alpha}^{}.
\label{eq:inverse_normal_mode_transform_1d}
\end{equation}

A convenient real Fourier basis is
\begin{align}
U_{s1}^{}&=\frac{1}{\sqrt{P}},
\nonumber\\
U_{s,2k}^{}&=\sqrt{\frac{2}{P}}
\cos\!\left[\frac{2\pi k(s-1)}{P}\right],
\nonumber\\
U_{s,2k+1}^{}&=\sqrt{\frac{2}{P}}
\sin\!\left[\frac{2\pi k(s-1)}{P}\right],
\label{eq:U_real_fourier_basis}
\end{align}
for \(k=1,\ldots,\lfloor(P-1)/2\rfloor\).  For even \(P\),
\begin{equation}
U_{sP}^{}=\frac{1}{\sqrt{P}}(-1)^{s-1}.
\label{eq:U_evenP_definition}
\end{equation}
With Eq.~\eqref{eq:PA_diagonalization},
\begin{equation}
\lambda_1^{}=0,
\quad
\lambda_{2k}^{}=\lambda_{2k+1}^{}
=4P^{}\sin^2\!\left(\frac{\pi k}{P}\right),
\label{eq:lambda_cos_sin_modes_1d}
\end{equation}
with \(\lambda_P=4P\) for the unpaired mode when \(P\) is even.  Thus
\(\lambda_\alpha\) is an eigenvalue of \(P\mathbf{A}\).

The Jacobian is constant:
\begin{equation}
J_P^{}
\equiv
\left|\det\frac{\partial \mathbf{x}}{\partial \mathbf{q}}\right|
=\left|\det(\sqrt{P}\,\mathbf{U})\right|
=P^{P/2},
\quad d\mathbf{x}=J_P^{}d\mathbf{q}.
\label{eq:normal_mode_jacobian_definition}
\end{equation}
Using Eqs.~\eqref{eq:PA_diagonalization} and
\eqref{eq:normal_mode_transform_1d},
\begin{equation}
V_{\rm eff}^{}(\mathbf{q})
=\sum_{\alpha=2}^{P}\frac{1}{2}k_{\alpha}^{}q_{\alpha}^{2}
+\phi(\mathbf{q}),
\label{eq:Veff_split_algo}
\end{equation}
where
\begin{equation}
k_{\alpha}^{}
\equiv
m^{}\omega_P^{2}\lambda_{\alpha}^{}
=\frac{m^{}P^{}\lambda_{\alpha}^{}}{\beta^{2}\hbar^{2}}
\label{eq:kalpha_definition}
\end{equation}
and
\begin{equation}
\phi(\mathbf{q})
\equiv
\frac{1}{P}\sum_{s=1}^{P}V(x_s^{}(\mathbf{q})).
\label{eq:phi_definition}
\end{equation}
Its normal-mode gradient is
\begin{equation}
\frac{\partial\phi}{\partial q_{\alpha}^{}}
=\frac{1}{\sqrt{P}}\sum_{s=1}^{P}U_{s\alpha}^{}V'(x_s^{}),
\label{eq:phi_force_transform}
\end{equation}
and in particular
\begin{equation}
\frac{\partial\phi}{\partial q_{\rm c}^{}}
=\frac{1}{P}\sum_{s=1}^{P}V'(x_s^{}).
\label{eq:phi_centroid_gradient}
\end{equation}
Throughout this paper, \(\phi\) denotes only the physical-potential part; the
spring contribution is kept explicitly in Eq.~\eqref{eq:Veff_split_algo}.

\subsection{Kubo-transformed correlations in normal modes}

We now rewrite Eq.~\eqref{eq:Kubo_path_QDO_final_rp} in normal-mode variables
to prepare for the BCMD dynamical approximation.  For brevity, throughout
this subsection we write
\begin{equation}
V_{\rm eff}^{}(\mathbf{q})\equiv V_{\rm eff}^{\,{\rm rp}}(\mathbf{q}).
\label{eq:Veff_abbreviation_bcmd}
\end{equation}
For a position-dependent operator \(\hat A^{}=A(\hat x^{})\), we define
\begin{equation}
\bar A_P^{\,{\rm nm}}(\mathbf{q})
\equiv
\frac{1}{P}\sum_{s=1}^{P}A\!\left(x_s^{}(\mathbf{q})\right),
\label{eq:A_P_nm_def}
\end{equation}
and, for \(\hat B^{}=B(\hat x^{})\),
\begin{equation}
\bar B_P^{\,{\rm nm}}(\mathbf{q})
\equiv
\frac{1}{P}\sum_{s=1}^{P}B\!\left(x_s^{}(\mathbf{q})\right).
\label{eq:B_P_nm_def}
\end{equation}

The configurational density transforms with the constant Jacobian as
\begin{equation}
\bar\rho_{\beta,P}^{\,{\rm nm}}(\mathbf{q})
\equiv
J_P^{}\,
\bar\rho_{\beta,P}^{\,{\rm rp}}(\mathbf{x}(\mathbf{q})),
\quad
\bar\rho_{\beta,P}^{\,{\rm nm}}(\mathbf{q})\,d\mathbf{q}
=
\bar\rho_{\beta,P}^{\,{\rm rp}}(\mathbf{x})\,d\mathbf{x}.
\label{eq:rho_nm_from_rp}
\end{equation}
Accordingly, the full extended equilibrium weight in normal-mode variables is
\begin{align}
W_{\beta,P}^{}(p_{\rm c}^{},\mathbf{q})
\equiv
J_P^{}\,
\rho_{\beta,P}^{\,{\rm rp}}(\mathbf{x}(\mathbf{q}),p_{\rm c}^{})
=
M_{\beta}^{}(p_{\rm c}^{})
\bar\rho_{\beta,P}^{\,{\rm nm}}(\mathbf{q}).
\label{eq:WBCMD}
\end{align}
Thus
\begin{equation}
Z_P^{}
=
\int d\mathbf{q}\,\bar\rho_{\beta,P}^{\,{\rm nm}}(\mathbf{q})
=
\int d\mathbf{q}\int dp_{\rm c}^{}\,
W_{\beta,P}^{}(p_{\rm c}^{},\mathbf{q}),
\label{eq:ZP_W_definition}
\end{equation}
in agreement with Eq.~\eqref{eq:ZP_rp_definition_path_QDO}.
Thus \(W_{\beta,P}(p_{\rm c},\mathbf q)\) is the same extended
equilibrium weight as \(\rho_{\beta,P}^{\rm rp}(\mathbf x,p_{\rm c})\),
expressed in normal-mode coordinates and including the constant Jacobian
\(J_P\).
The factorization of \(W_{\beta,P}\) follows directly from the ring-polymer representation in
Eq.~\eqref{eq:rho_rp_pc_factorization_early}; it is not an additional 
assumption.

The same Jacobian transforms the unnormalized dynamical object:
\begin{equation}
\widetilde B_{\beta,P}^{\,{\rm nm}}(p_{\rm c}^{},\mathbf{q},t)
\equiv
J_P^{}\,
\widetilde B_{\beta,P}^{\,{\rm rp}}
(\mathbf{x}(\mathbf{q}),p_{\rm c}^{},t).
\label{eq:Btilde_nm_definition}
\end{equation}
Equivalently,
\begin{align}
\widetilde B_{\beta,P}^{\,{\rm nm}}(p_{\rm c}^{},\mathbf{q},t)
\,d\mathbf{q}\,dp_{\rm c}^{}
=
\widetilde B_{\beta,P}^{\,{\rm rp}}(\mathbf{x},p_{\rm c}^{},t)
\,d\mathbf{x}\,dp_{\rm c}^{}.
\label{eq:Btilde_nm_measure_transform}
\end{align}

The normalized normal-mode estimator is
\begin{equation}
\bar B_{\beta,P}^{\,{\rm nm}}(p_{\rm c}^{},\mathbf{q};t)
\equiv
\frac{
\widetilde B_{\beta,P}^{\,{\rm nm}}(p_{\rm c}^{},\mathbf{q},t)
}{
W_{\beta,P}^{}(p_{\rm c}^{},\mathbf{q})
}.
\label{eq:Bbar_nm_definition}
\end{equation}
Because the numerator and denominator transform with the same Jacobian,
\begin{equation}
\bar B_{\beta,P}^{\,{\rm nm}}(p_{\rm c}^{},\mathbf{q};t)
=
\bar B_{\beta,P}^{\,{\rm rp}}
(\mathbf{x}(\mathbf{q}),p_{\rm c}^{};t).
\label{eq:Bbar_nm_coordinate_invariance}
\end{equation}
In particular, for a position-dependent \(\hat B^{}\),
\begin{equation}
\bar B_{\beta,P}^{\,{\rm nm}}(p_{\rm c}^{},\mathbf{q};0)
=
\bar B_P^{\,{\rm nm}}(\mathbf{q}),
\label{eq:Bbar_nm_initial}
\end{equation}
which is independent of \(p_{\rm c}^{}\).

Changing variables in Eq.~\eqref{eq:Kubo_path_QDO_final_rp} now gives
\begin{align}
C_{AB}^{\,{\rm K}}(t)
&=
\lim_{P\to\infty}
\frac{1}{Z_P^{}}
\int d\mathbf{q}\int dp_{\rm c}^{}\,
W_{\beta,P}^{}(p_{\rm c}^{},\mathbf{q})
\nonumber\\
&\quad\times
\bar A_P^{\,{\rm nm}}(\mathbf{q})
\bar B_{\beta,P}^{\,{\rm nm}}(p_{\rm c}^{},\mathbf{q};t).
\label{eq:exact_Kubo_nm_W_form}
\end{align}
All real-time quantum dynamics is contained in the final estimator.  At a
fixed \(P\), its dependence on the initial position-diagonal observable may be denoted
formally by
\begin{equation}
\bar B_{\beta,P}^{\,{\rm nm}}(p_{\rm c}^{},\mathbf{q};t)
=
e^{t\mathcal K_{\rm ex}^{}(t)}\bar B_P^{\,{\rm nm}}(\mathbf{q}).
\label{eq:Kex_formal_nm_pc}
\end{equation}
Here \(\mathcal K_{\rm ex}^{}(t)\) is a formal time-dependent logarithmic
generator, defined so that \(e^{t\mathcal K_{\rm ex}(t)}\) reproduces the
exact reduced propagation induced by the path-QDO representation and the exact
Heisenberg evolution of the position-diagonal observable \(\hat B\).  Equivalently, whenever the operator logarithm
is defined, \(t\mathcal K_{\rm ex}(t)\) is the logarithm of that exact reduced
propagation.  No Markov or semigroup property is implied; in general
\(\mathcal K_{\rm ex}(t)\) may depend on \(t\).  Equation~\eqref{eq:Kex_formal_nm_pc}
\begingroup\color{black}
therefore serves only as a formal representation of the exact reduced
dynamics, without introducing an independent closure assumption. In particular,
no explicit path-QDO Liouvillian is derived here.
\endgroup

For later use, we define the normalized finite-\(P\) average
\begin{equation}
\langle f\rangle_W
\equiv
\frac{1}{Z_P^{}}
\int d\mathbf{q}\int dp_{\rm c}^{}\,
W_{\beta,P}^{}(p_{\rm c}^{},\mathbf{q})
f(p_{\rm c}^{},\mathbf{q}).
\label{eq:W_one_point_average}
\end{equation}
The statistical invariance of the exact dynamics can then be stated in the
normal-mode representation.  Setting \(\bar A_P^{\,{\rm nm}}=1\) in
Eq.~\eqref{eq:exact_Kubo_nm_W_form} gives, for a position-dependent
\(\hat B=B(\hat x)\),
\begin{equation}
\lim_{P\to\infty}
\left\langle
e^{t \mathcal K_{\rm ex}^{}(t)}\bar B_P^{\,{\rm nm}}
\right\rangle_W
=
\langle\hat B^{}\rangle
=
\lim_{P\to\infty}
\left\langle
\bar B_P^{\,{\rm nm}}
\right\rangle_W.
\label{eq:Kex_preserves_static_average}
\end{equation}
Thus \(e^{t\mathcal K_{\rm ex}^{}(t)}\) changes the path-resolved estimator while
preserving its equilibrium average in the continuum imaginary-time limit.
This relation expresses the stationarity and time-translation symmetry
inherited from the exact Heisenberg evolution.

\subsection{BCMD generator}

The exact reduced propagation derived above is formally exact but not
computationally useful: evaluating \(\mathcal K_{\rm ex}(t)\) requires the exact
Heisenberg dynamics and, after reduction to the path variables, need not yield
a closed Markovian equation.  BCMD therefore introduces a dynamical closure by replacing
\(e^{t\mathcal K_{\rm ex}(t)}\) with a time-homogeneous Markov propagator
\(e^{t\mathcal K_{\rm bcmd}}\).  This replacement is not a formal
consequence of the path-QDO closure.  In particular, the path-QDO
provides the exact path-resolved reference framework on which the BCMD
approximation is formulated, but it does not uniquely select or derive
\(\mathcal K_{\rm bcmd}\); the latter is an additional dynamical approximation.
Instead, \(\mathcal K_{\rm bcmd}\)
is constructed to preserve the finite-\(P\) equilibrium weight
\(W_{\beta,P}\), recover the exact equal-time Kubo value,
reduce to classical Hamiltonian dynamics for \(P=1\), and retain the exact
centroid dynamics for a harmonic system with a linear position observable.

The centroid and non-centroid modes are assigned different dynamical roles.
The path-QDO retains the physical conjugate pair \((q_{\rm c},p_{\rm c})\), so
the centroid is propagated reversibly with the physical mass.  The
non-centroid coordinates instead describe fluctuations of the imaginary-time
path and have no corresponding physical momenta in the reduced representation.
Assigning them inertial Hamiltonian dynamics would therefore introduce
arbitrary dynamical frequencies.  BCMD treats these modes instead as Brownian
variables whose dynamics preserves the ring-polymer equilibrium distribution,
allowing the path fluctuations to relax without interpreting their normal-mode
frequencies as physical vibrations.

BCMD is consequently expected to be most reliable when the real-time response
is governed mainly by centroid or collective physical motion, while the
non-centroid modes primarily represent equilibrium quantum fluctuations.
Coherent dynamics carried specifically by the non-centroid modes lies outside
this physical picture.  The resulting closure is therefore approximate but
equilibrium preserving and computationally tractable.

The extended BCMD state is given by
\begin{equation}
\mathbf{y}
\equiv
\left(p_{\rm c}^{},q_{\rm c}^{},q_2^{},\ldots,q_P^{}\right)
\equiv
\left(p_{\rm c}^{},\mathbf{q}\right),
\label{eq:y_state}
\end{equation}
and the dynamical approximation is
\begin{equation}
\bar B_{\beta,P}^{\,{\rm nm}}(p_{\rm c}^{},\mathbf{q};t)
\approx
\left[
 e^{t\mathcal K_{\rm bcmd}^{}}
 \bar B_P^{\,{\rm nm}}
\right](p_{\rm c}^{},\mathbf{q}).
\label{eq:BCMD_propagator_replacement}
\end{equation}
The backward generator is chosen as
\begin{equation}
\mathcal K_{\rm bcmd}^{}
\equiv
\mathcal K_{\rm c}^{}+\mathcal K_{\rm nc}^{}.
\label{eq:KBCMD_sum}
\end{equation}
The centroid part is the Hamiltonian Liouville operator
\begin{equation}
\mathcal K_{\rm c}^{}
\equiv
\frac{p_{\rm c}^{}}{m^{}}
\frac{\partial}{\partial q_{\rm c}^{}}
-
\frac{\partial V_{\rm eff}^{}}{\partial q_{\rm c}^{}}
\frac{\partial}{\partial p_{\rm c}^{}}.
\label{eq:Kc}
\end{equation}
Since the spring term contains no centroid coordinate,
\begin{equation}
\frac{\partial V_{\rm eff}^{}}{\partial q_{\rm c}^{}}
=
\frac{\partial\phi}{\partial q_{\rm c}^{}}.
\label{eq:centroid_force_phi}
\end{equation}
The non-centroid part is the backward Smoluchowski operator
\begin{equation}
\mathcal K_{\rm nc}^{}
\equiv
\sum_{\alpha=2}^{P}
\frac{1}{\beta^{}m^{}\gamma_{\alpha}^{}}
 e^{\beta^{}V_{\rm eff}^{}}
\frac{\partial}{\partial q_{\alpha}^{}}
\left[
 e^{-\beta^{}V_{\rm eff}^{}}
\frac{\partial}{\partial q_{\alpha}^{}}
\right],
\label{eq:Knc}
\end{equation}
or, equivalently,
\begin{equation}
\mathcal K_{\rm nc}^{}
=
\sum_{\alpha=2}^{P}
\left[
-
\frac{1}{m^{}\gamma_{\alpha}^{}}
\frac{\partial V_{\rm eff}^{}}{\partial q_{\alpha}^{}}
\frac{\partial}{\partial q_{\alpha}^{}}
+
\frac{1}{\beta^{}m^{}\gamma_{\alpha}^{}}
\frac{\partial^{2}}{\partial q_{\alpha}^{2}}
\right].
\label{eq:Knc_expanded}
\end{equation}

Let \(\mathcal K^\dagger\) denote the forward adjoint of \(\mathcal K\).  Using
\(\partial_{p_{\rm c}}W_{\beta,P}=-(\beta p_{\rm c}/m)W_{\beta,P}\) and
\(\partial_{q_{\alpha}}W_{\beta,P}= -\beta
(\partial_{q_{\alpha}}V_{\rm eff})W_{\beta,P}\), one finds
\begin{align}
\mathcal K_{\rm c}^{\dagger}W_{\beta,P}^{}
&=
-\frac{\partial}{\partial q_{\rm c}^{}}
\left(\frac{p_{\rm c}^{}}{m^{}}W_{\beta,P}^{}\right)
+
\frac{\partial}{\partial p_{\rm c}^{}}
\left(
\frac{\partial V_{\rm eff}^{}}{\partial q_{\rm c}^{}}
W_{\beta,P}^{}
\right)
=0,
\label{eq:Kc_stationarity_explicit}
\\
\mathcal K_{\rm nc}^{\dagger}W_{\beta,P}^{}
&=
\sum_{\alpha=2}^{P}
\frac{1}{m^{}\gamma_{\alpha}^{}}
\frac{\partial}{\partial q_{\alpha}^{}}
\left[
\frac{\partial V_{\rm eff}^{}}{\partial q_{\alpha}^{}}
W_{\beta,P}^{}
+
\frac{1}{\beta^{}}
\frac{\partial W_{\beta,P}^{}}{\partial q_{\alpha}^{}}
\right]
=0.
\label{eq:Knc_stationarity_explicit}
\end{align}
Therefore
\begin{equation}
\mathcal K_{\rm bcmd}^{\dagger}W_{\beta,P}^{}=0,
\label{eq:KBCMD_stationary}
\end{equation}
which leads to
\begin{equation}
\left\langle e^{t\mathcal K_{\rm bcmd}^{}}f\right\rangle_W
=
\langle f\rangle_W.
\label{eq:Kbcmd_preserves_W_average}
\end{equation}
In particular, for the bead estimator of a position-dependent operator,
\begin{equation}
\lim_{P\to\infty}
\left\langle
 e^{t\mathcal K_{\rm bcmd}^{}}\bar B_P^{\,{\rm nm}}
\right\rangle_W
=
\lim_{P\to\infty}
\left\langle\bar B_P^{\,{\rm nm}}\right\rangle_W
=
\langle\hat B^{}\rangle ,
\label{eq:Kbcmd_preserves_quantum_static_average}
\end{equation}
which verifies explicitly the equilibrium-preservation condition.

The stochastic differential equations corresponding to this backward
generator are
\begin{align}
\dot q_{\rm c}^{}
&=
\frac{p_{\rm c}^{}}{m^{}},
\label{eq:bcmd_centroid_eqn_a}
\\
\dot p_{\rm c}^{}
&=
-\frac{\partial V_{\rm eff}^{}}{\partial q_{\rm c}^{}}
=
-\frac{\partial\phi}{\partial q_{\rm c}^{}} ,
\label{eq:bcmd_centroid_eqn_b}
\end{align}
and, for \(\alpha=2,\ldots,P\),
\begin{equation}
m^{}\gamma_{\alpha}^{}\dot q_{\alpha}^{}
=
-\frac{\partial V_{\rm eff}^{}}{\partial q_{\alpha}^{}}
+
\sqrt{\frac{2m^{}\gamma_{\alpha}^{}}{\beta^{}}}\,
\zeta_{\alpha}^{}(t),
\label{eq:bcmd_noncentroid_eqn}
\end{equation}
where
\begin{equation}
\left\langle
\zeta_{\alpha}^{}(t)\zeta_{\alpha'}^{}(t')
\right\rangle
=
\delta_{\alpha\alpha'}^{}\delta(t-t').
\label{eq:bcmd_noise}
\end{equation}
Using
\begin{equation}
V_{\rm eff}^{}(\mathbf{q})
=
\frac{1}{2}\sum_{\alpha=2}^{P}k_{\alpha}^{}q_{\alpha}^{2}
+\phi(\mathbf{q}),
\label{eq:Veff_split_bcmd}
\end{equation}
Eq.~\eqref{eq:bcmd_noncentroid_eqn} becomes
\begin{equation}
m^{}\gamma_{\alpha}^{}\dot q_{\alpha}^{}
=
-k_{\alpha}^{}q_{\alpha}^{}
-\frac{\partial\phi}{\partial q_{\alpha}^{}}
+
\sqrt{\frac{2m^{}\gamma_{\alpha}^{}}{\beta^{}}}\,
\zeta_{\alpha}^{}(t).
\label{eq:bcmd_noncentroid_split_form}
\end{equation}
This is the form used in the numerical algorithm below.

For a free ring polymer, \(\phi=0\), each non-centroid mode is an overdamped
Ornstein--Uhlenbeck process,
\begin{equation}
m^{}\gamma_{\alpha}^{}\dot q_{\alpha}^{}
=
-k_{\alpha}^{}q_{\alpha}^{}
+
\sqrt{\frac{2m^{}\gamma_{\alpha}^{}}{\beta^{}}}\,
\zeta_{\alpha}^{}(t).
\label{eq:free_rp_potential}
\end{equation}
Its relaxation time is
\begin{equation}
\tau_{\alpha}^{}
=
\frac{m^{}\gamma_{\alpha}^{}}{k_{\alpha}^{}}
=
\frac{\beta^{2}\hbar^{2}\gamma_{\alpha}^{}}
{P\lambda_{\alpha}^{}}.
\label{eq:tau_alpha}
\end{equation}
Following the original BCMD construction,\cite{Shiga2022BCMD} we choose the
same relaxation time for all non-centroid modes,
\begin{equation}
\tau_{\alpha}^{}=\beta^{}\hbar,
\quad \alpha=2,\ldots,P.
\label{eq:tau_alpha_bhbar}
\end{equation}
The motivation is to relate the relaxation time assigned to the
imaginary-time path fluctuations to the thermal quantum time scale rather
than to any ring-polymer normal-mode frequency.  In the free-particle
argument of Ref.~\onlinecite{Shiga2022BCMD}, a Gaussian wave packet whose
momentum probability distribution is Maxwell--Boltzmann has the initial
spatial standard deviation
\(\sigma_x(0)=\sqrt{\beta\hbar^2/(4m)}\)
and spreads on the time scale of \(\beta\hbar\): its width changes
appreciably by \(t\sim\beta\hbar/2\) and is larger by a factor of
\(\sqrt{5}\) at \(t=\beta\hbar\).  Notably, this initial spatial standard
deviation coincides, for a free particle, with the root-mean-square
separation between two points of the imaginary-time path that are half an
imaginary-time period apart, where the separation is maximal.  This
correspondence further supports a heuristic physical link between the
free-particle wave-packet spreading and the relaxation of the
imaginary-time path fluctuations.  BCMD therefore assigns the same
relaxation time \(\beta\hbar\) to all non-centroid modes, rather than tying
their relaxation times to the ring-polymer normal-mode frequencies.

With the convention
\(k_{\alpha}=mP\lambda_{\alpha}/(\beta^{2}\hbar^{2})\) of
Eq.~\eqref{eq:kalpha_definition}, Eq.~\eqref{eq:tau_alpha_bhbar} is enforced by
\begin{equation}
\gamma_{\alpha}^{}
=
\frac{P\lambda_{\alpha}^{}}{\beta^{}\hbar}.
\label{eq:gamma_choice}
\end{equation}
Here \(\lambda_{\alpha}\) is the eigenvalue of \(P\mathbf{A}\), as defined in
Eq.~\eqref{eq:PA_diagonalization}.  Equations~\eqref{eq:kalpha_definition} and
\eqref{eq:gamma_choice} therefore reproduce the normal-mode and relaxation-time
convention of the original BCMD formulation.\cite{Shiga2022BCMD}

Within BCMD,
\begin{equation}
C_{AB}^{\,{\rm K}}(t)
\approx
\lim_{P\to\infty}C_{AB,P}^{\,{\rm bcmd}}(t),
\label{eq:BCMD_Kubo_approx}
\end{equation}
where
\begin{align}
C_{AB,P}^{\,{\rm bcmd}}(t)
&\equiv
\frac{1}{Z_P^{}}
\int d\mathbf{q}\int dp_{\rm c}^{}\,
W_{\beta,P}^{}(p_{\rm c}^{},\mathbf{q})
\bar A_P^{\,{\rm nm}}(\mathbf{q})
\nonumber\\
&\quad\times
\left[
 e^{t\mathcal K_{\rm bcmd}^{}}
 \bar B_P^{\,{\rm nm}}
\right](p_{\rm c}^{},\mathbf{q}).
\label{eq:BCMD_Kubo_formula}
\end{align}
For compactness, we define
\begin{equation}
\langle f,g\rangle_W
\equiv
\frac{1}{Z_P^{}}
\int d\mathbf{q}\int dp_{\rm c}^{}\,
W_{\beta,P}^{}(p_{\rm c}^{},\mathbf{q})
 f(p_{\rm c}^{},\mathbf{q})g(p_{\rm c}^{},\mathbf{q}).
\label{eq:BCMD_weighted_inner_product}
\end{equation}
Then
\begin{equation}
C_{AB,P}^{\,{\rm bcmd}}(t)
=
\left\langle
\bar A_P^{\,{\rm nm}},
 e^{t\mathcal K_{\rm bcmd}^{}}\bar B_P^{\,{\rm nm}}
\right\rangle_W.
\label{eq:CAB_BCMD_inner_product_form}
\end{equation}

At \(t=0\), the dynamical propagator is the identity and the normalized
Maxwell factor integrates to unity.  Therefore, for position-dependent
\(\hat A^{}\) and \(\hat B^{}\),
\begin{equation}
C_{AB,P}^{\,{\rm bcmd}}(0)
=
\frac{1}{Z_P^{}}
\int d\mathbf{q}\,
\bar\rho_{\beta,P}^{\,{\rm nm}}(\mathbf{q})
\bar A_P^{\,{\rm nm}}(\mathbf{q})
\bar B_P^{\,{\rm nm}}(\mathbf{q}).
\label{eq:CAB_BCMD_t0_finiteP}
\end{equation}
Taking the continuum imaginary-time limit gives
\begin{equation}
\lim_{P\to\infty}C_{AB,P}^{\,{\rm bcmd}}(0)
=
C_{AB}^{\,{\rm K}}(0).
\label{eq:CAB_BCMD_t0_exact_limit_rewrite}
\end{equation}
Thus the equal-time Kubo value is recovered exactly for position-diagonal
operators, while the BCMD approximation affects only the subsequent real-time
propagation.



\section{Basic properties of BCMD} \label{subsec:bcmd_properties}

\begingroup\color{black}
The equilibrium, classical-limit, time-symmetry, short-time, and harmonic-oscillator
properties of BCMD were established in Ref.~\onlinecite{Shiga2022BCMD}.
These established properties are rederived here using the present BCMD generator
and normal-mode representation. In particular, time symmetry is shown using
its weighted adjoint, while the short-time coefficients follow from powers
of the generator. The equal-time results are also connected to the path-QDO
representation of Sec.~II.
\endgroup
\subsection{Thermal equilibrium}

As shown in Eqs.~\eqref{eq:KBCMD_stationary} and
\eqref{eq:Kbcmd_preserves_W_average}, the BCMD generator leaves the
finite-\(P\) equilibrium weight \(W_{\beta,P}\) stationary and preserves all
averages evaluated with that weight.  For the bead estimators of
position-dependent observables, these averages converge to the corresponding
quantum canonical averages as \(P\to\infty\), as stated in
Eq.~\eqref{eq:Kbcmd_preserves_quantum_static_average}.

\subsection{Classical limit}

When \(P=1\), there is no non-centroid mode,
so \(\mathcal K_{\rm nc}^{}\) is absent,
\(q_{\rm c}^{}=x^{}\),
\(p_{\rm c}^{}=p^{}\),
and \(V_{\rm eff}^{}=\phi=V(x)\).
Therefore
\begin{equation}
\mathcal K_{\rm bcmd}^{}
\to
\frac{p^{}}{m^{}} \frac{\partial}{\partial x^{}}
-
\frac{\partial V}{\partial x^{}} \frac{\partial}{\partial p^{}},
\label{eq:Kbcmd_classical_limit_rewrite}
\end{equation}
namely the classical Liouville operator.
Equation~\eqref{eq:CAB_BCMD_inner_product_form} then reduces to the
classical canonical correlation function.
Hence BCMD has the correct classical limit.

\subsection{Time symmetry}

Let \(\Theta\) denote centroid-momentum reversal,
\begin{equation}
(\Theta f)(p_{\rm c}^{},\mathbf{q})
\equiv
f(-p_{\rm c}^{},\mathbf{q}) .
\label{eq:Theta_definition_rewrite}
\end{equation}
Let \(\mathcal K^{\ddagger}\) denote the adjoint with respect to the weighted inner
product in Eq.~\eqref{eq:BCMD_weighted_inner_product}, such that
\begin{equation}
\left\langle f,\mathcal K g\right\rangle_W
=
\left\langle \mathcal K^{\ddagger} f,g\right\rangle_W .
\label{eq:weighted_adjoint_definition}
\end{equation}
This weighted adjoint is distinct from the ordinary forward adjoint
\(\mathcal K^\dagger\) used in the equilibrium-preservation discussion; the
two are related by
\begin{equation}
\mathcal K^{\ddagger}f
=
W_{\beta,P}^{-1}\mathcal K^\dagger\!\left(W_{\beta,P}f\right).
\label{eq:weighted_adjoint_relation}
\end{equation}
The centroid Liouvillian is antisymmetric
in the weighted inner product, whereas the non-centroid Smoluchowski operator
is symmetric.  Together with momentum reversal, this gives
\begin{equation}
\Theta \mathcal K_{\rm c}^{} \Theta = -\mathcal K_{\rm c}^{} = \mathcal K_{\rm c}^{\ddagger},
\quad
\Theta \mathcal K_{\rm nc}^{} \Theta = \mathcal K_{\rm nc}^{} = \mathcal K_{\rm nc}^{\ddagger},
\label{eq:Theta_K_parts_rewrite}
\end{equation}
and hence
\begin{equation}
\Theta \mathcal K_{\rm bcmd}^{} \Theta = \mathcal K_{\rm bcmd}^{\ddagger} .
\label{eq:Theta_Kbcmd_rewrite}
\end{equation}
Using \(\Theta^2=1\), one has
\begin{equation}
\Theta \mathcal K_{\rm bcmd}^{n}\Theta
=
\left(\Theta\mathcal K_{\rm bcmd}\Theta\right)^n,
\label{eq:Theta_K_power_relation}
\end{equation}
and thus, exponentiating Eq.~\eqref{eq:Theta_Kbcmd_rewrite},
\begin{equation}
\Theta e^{t\mathcal K_{\rm bcmd}}\Theta
=
e^{t\mathcal K_{\rm bcmd}^{\ddagger}}.
\label{eq:Theta_propagator_relation}
\end{equation}
By the definition of the weighted adjoint,
\begin{equation}
\left\langle f,e^{t\mathcal K_{\rm bcmd}}g\right\rangle_W
=
\left\langle e^{t\mathcal K_{\rm bcmd}^{\ddagger}}f,g\right\rangle_W.
\label{eq:weighted_propagator_adjoint}
\end{equation}
Using these relations together with the invariance of
\(W_{\beta,P}^{}\) under \(p_{\rm c}^{}\mapsto -p_{\rm c}^{}\), one obtains
\begin{equation}
\left\langle
f, e^{t\mathcal K_{\rm bcmd}^{}} g
\right\rangle_{W}
=
\left\langle
\Theta g, e^{t\mathcal K_{\rm bcmd}^{}} \Theta f
\right\rangle_{W} .
\label{eq:BCMD_reversibility_identity}
\end{equation}
This is the same formal reversibility structure that underlies the
usual symmetry discussion for CMD and RPMD.

For position-diagonal observables,
\(\bar A_P^{\,{\rm nm}}\) and \(\bar B_P^{\,{\rm nm}}\) are even under
\(\Theta\), and therefore
\begin{equation}
C_{AB,P}^{\,{\rm bcmd}}(t)
=
C_{BA,P}^{\,{\rm bcmd}}(t) .
\label{eq:CAB_BCMD_symmetry_same_t}
\end{equation}
Equivalently, extending the correlation to negative times by
\begin{equation}
C_{AB,P}^{\,{\rm bcmd}}(-t)
\equiv
C_{BA,P}^{\,{\rm bcmd}}(t) ,
\quad
t>0 ,
\label{eq:CAB_negative_time_extension}
\end{equation}
then one has the same formal symmetry as for the exact Kubo transform,
\begin{equation}
C_{AB,P}^{\,{\rm bcmd}}(t)
=
C_{BA,P}^{\,{\rm bcmd}}(-t) .
\label{eq:CAB_BCMD_Kubo_like_symmetry}
\end{equation}
In particular, for an autocorrelation function,
\begin{equation}
C_{AA,P}^{\,{\rm bcmd}}(t)
=
C_{AA,P}^{\,{\rm bcmd}}(-t) ,
\label{eq:CAA_even_time_rewrite}
\end{equation}
so it is an even real function of time.

\subsection{Spectra}

The Fourier transform is defined by
\begin{equation}
\tilde C_{AB,P}^{\,{\rm bcmd}}(\omega)
\equiv
\int_{-\infty}^{\infty} dt\,
e^{i\omega t}
C_{AB,P}^{\,{\rm bcmd}}(t) .
\label{eq:CAB_spectrum_definition_rewrite}
\end{equation}
For an autocorrelation of a real position-diagonal observable,
Eq.~\eqref{eq:CAA_even_time_rewrite} implies
\begin{equation}
\tilde C_{AA,P}^{\,{\rm bcmd}}(\omega)
\in \mathbb{R},
\quad
\tilde C_{AA,P}^{\,{\rm bcmd}}(-\omega)
=
\tilde C_{AA,P}^{\,{\rm bcmd}}(\omega) .
\label{eq:Caa_spectrum_real_even_rewrite}
\end{equation}
Thus the BCMD spectrum has the same basic reality and evenness
properties as the exact Kubo spectrum.

The connection between Kubo-transformed correlation functions and spectral
sum rules has been discussed previously.\cite{BraamsMillerManolopoulos2006}
For the present normalization, the zeroth-moment sum rule follows immediately:
\begin{equation}
\int_{-\infty}^{\infty}
\frac{d\omega}{2\pi}\,
\tilde C_{AA,P}^{\,{\rm bcmd}}(\omega)
=
C_{AA,P}^{\,{\rm bcmd}}(0) .
\label{eq:zeroth_moment_sum_rule_rewrite}
\end{equation}
Since the equal-time value is exact in the fine-discretization limit
for position-diagonal operators, the corresponding zeroth moment is also
exact in that limit.

For the position autocorrelation, the estimator is simply the centroid
coordinate:
\begin{equation}
\bar x_P^{\,{\rm nm}}(\mathbf{q})
=
\frac{1}{P}\sum_{s=1}^{P} x_s^{}(\mathbf{q})
=
q_{\rm c}^{} .
\label{eq:x_centroid_rewrite}
\end{equation}
Therefore
\begin{equation}
C_{xx,P}^{\,{\rm bcmd}}(t)
=
\left\langle
q_{\rm c}^{},
e^{t\mathcal K_{\rm bcmd}^{}} q_{\rm c}^{}
\right\rangle_{W} .
\label{eq:Cxx_BCMD_rewrite}
\end{equation}
Since
\begin{equation}
\mathcal K_{\rm bcmd}^{} q_{\rm c}^{}
=
\frac{p_{\rm c}^{}}{m^{}} \equiv v_{\rm c}^{} ,
\label{eq:K_on_qc_rewrite}
\end{equation}
it is natural to define the associated velocity correlation by
\begin{equation}
C_{vv,P}^{\,{\rm bcmd}}(t)
\equiv
-
\frac{d^{2}}{dt^{2}} C_{xx,P}^{\,{\rm bcmd}}(t) .
\label{eq:Cvv_from_Cxx_rewrite}
\end{equation}
Using the generalized detailed-balance identity derived above, the same
quantity can equivalently be written as the centroid-velocity correlation
\begin{equation}
C_{vv,P}^{\,{\rm bcmd}}(t)
=
\left\langle
v_{\rm c}^{},
e^{t\mathcal K_{\rm bcmd}^{}}v_{\rm c}^{}
\right\rangle_W .
\label{eq:Cvv_direct_rewrite}
\end{equation}
Its Fourier transform then satisfies
\begin{equation}
\tilde C_{vv,P}^{\,{\rm bcmd}}(\omega)
=
\omega^{2}
\tilde C_{xx,P}^{\,{\rm bcmd}}(\omega) .
\label{eq:Cvv_spectrum_from_Cxx_rewrite}
\end{equation}

For the $3N$-dimensional extension, let
\(v_{k\alpha,{\rm c}}^{}\) denote the centroid velocity of Cartesian component
\(\alpha\) of atom \(k\), with mass \(m_k^{}\).
We define the BCMD vibrational density of states by
\begin{equation}
D_{P}^{\,{\rm bcmd}}(\omega)
\equiv
\beta^{}
\sum_{k=1}^{N}
\sum_{\alpha=1}^{3}
m_k^{}
\tilde C_{v_{k\alpha,{\rm c}}^{}v_{k\alpha,{\rm c}}^{},P}^{\rm bcmd}(\omega) .
\label{eq:VDOS_definition_rewrite}
\end{equation}
Then
\begin{align}
&\int_{-\infty}^{\infty}
\frac{d\omega}{2\pi}\,
D_{P}^{\,{\rm bcmd}}(\omega)
=
\beta^{}
\sum_{k=1}^{N}\sum_{\alpha=1}^{3}
m_k^{}
C_{v_{k\alpha,{\rm c}}^{}v_{k\alpha,{\rm c}}^{},P}^{\rm bcmd}(0)
\nonumber\\
&=
\beta^{}
\sum_{k=1}^{N}\sum_{\alpha=1}^{3}
m_k^{}
\left\langle
v_{k\alpha,{\rm c}}^{2}
\right\rangle_{W}
=
\beta^{}
\sum_{k=1}^{N}\sum_{\alpha=1}^{3}
\frac{1}{\beta^{}}
=
3N .
\label{eq:VDOS_sum_rule_3N_rewrite}
\end{align}
The last equality follows directly from the Gaussian centroid-momentum
distribution, i.e.\ from equipartition.
Thus the usual \(3N\) sum rule for the vibrational density of states is
recovered in BCMD.

\subsection{Short-time limit}

The exact short-time coefficients of Kubo-transformed correlation
functions provide a standard benchmark for path-integral dynamical
approximations.\cite{BraamsManolopoulos2006}
The short-time expansion follows directly from
Eq.~\eqref{eq:CAB_BCMD_inner_product_form}:
\begin{equation}
C_{AB,P}^{\,{\rm bcmd}}(t)
=
\sum_{n=0}^{\infty}
\frac{t^{n}}{n!}
\left\langle
\bar A_P^{\,{\rm nm}},
\mathcal K_{\rm bcmd}^{n} \bar B_P^{\,{\rm nm}}
\right\rangle_{W} .
\label{eq:CAB_short_time_expansion_rewrite}
\end{equation}
Hence, for a general pair \(A,B\),
\begin{align}
C_{AB,P}^{\,{\rm bcmd}}(t)
=&
C_{AB,P}^{\,{\rm bcmd}}(0)
+
t\,
\left\langle
\bar A_P^{\,{\rm nm}},
\mathcal K_{\rm bcmd}^{} \bar B_P^{\,{\rm nm}}
\right\rangle_{W}
\nonumber \\ +&
\frac{t^{2}}{2}
\left\langle
\bar A_P^{\,{\rm nm}},
\mathcal K_{\rm bcmd}^{2} \bar B_P^{\,{\rm nm}}
\right\rangle_{W}
+
O(t^{3}) .
\label{eq:CAB_short_time_general_rewrite}
\end{align}
For an autocorrelation of a time-even observable, the two-sided correlation
is even by Eq.~\eqref{eq:CAA_even_time_rewrite}.  Because the BCMD generator
contains a Smoluchowski part, however, this symmetry does not in general imply
analyticity at \(t=0\); the short-time expansion below is therefore understood
for \(t\to0^+\).

For the position autocorrelation, we have
\begin{equation}
\bar x_P^{\,{\rm nm}}(\mathbf{q}) = q_{\rm c}^{},
\quad
v_{\rm c}^{} \equiv \frac{p_{\rm c}^{}}{m^{}} ,
\quad
F_{\rm c}^{}(\mathbf{q})
\equiv
-
\frac{\partial \phi}{\partial q_{\rm c}^{}} .
\label{eq:xc_vc_Fc_def_rewrite}
\end{equation}
Since the centroid derivative of \(V_{\rm eff}^{}\) receives no
contribution from the spring term,
\begin{equation}
\frac{\partial V_{\rm eff}^{}}{\partial q_{\rm c}^{}} =
\frac{\partial \phi}{\partial q_{\rm c}^{}} ,
\label{eq:dVeff_dqc_rewrite}
\end{equation}
and therefore
\begin{equation}
F_{\rm c}^{}(\mathbf{q})
=
-
\frac{1}{P}
\sum_{s=1}^{P}
V'\!\left( x_s^{}(\mathbf{q}) \right) .
\label{eq:Fc_bead_average_force_rewrite}
\end{equation}
Moreover,
\begin{equation}
\mathcal K_{\rm bcmd}^{} q_{\rm c}^{} = v_{\rm c}^{},
\quad
\mathcal K_{\rm bcmd}^{} v_{\rm c}^{} = \frac{F_{\rm c}^{}}{m^{}} .
\label{eq:K_qc_K_vc_rewrite}
\end{equation}

The first derivative at the origin vanishes:
\begin{equation}
\left.
\frac{d}{dt}
C_{xx,P}^{\,{\rm bcmd}}(t)
\right|_{t=0}
=
\left\langle
q_{\rm c}^{},
v_{\rm c}^{}
\right\rangle_{W}
=0 ,
\label{eq:Cxx_first_derivative_zero_rewrite}
\end{equation}
because the equilibrium distribution is even in \(p_{\rm c}^{}\).
Also,
\begin{equation}
C_{vv,P}^{\,{\rm bcmd}}(0)
=
\left\langle
v_{\rm c}^{2}
\right\rangle_{W}
=
\frac{1}{\beta^{} m^{}} ,
\label{eq:Cvv_t0_equipartition_rewrite}
\end{equation}
again by equipartition of the centroid momentum.

Using Eq.~\eqref{eq:Cvv_from_Cxx_rewrite}, one obtains
\begin{align}
C_{xx,P}^{\,{\rm bcmd}}(t)
&=
C_{xx,P}^{\,{\rm bcmd}}(0)
-
\frac{t^{2}}{2}
C_{vv,P}^{\,{\rm bcmd}}(0)
\nonumber\\
&\quad+
\frac{t^{4}}{24}
C_{xx,P}^{(4)}(0)
+
O(t^{5}) .
\label{eq:Cxx_short_time_general_even_rewrite}
\end{align}
The \(t^{2}\) coefficient is therefore
\begin{equation}
C_{xx,P}^{\,{\rm bcmd}}(t)
=
C_{xx,P}^{\,{\rm bcmd}}(0)
-
\frac{t^{2}}{2\beta^{}m^{}}
+
O(t^{4}) .
\label{eq:Cxx_short_time_t2_rewrite}
\end{equation}

To obtain the next term, note first that
\begin{equation}
\left.
\frac{d}{dt}
C_{vv,P}^{\,{\rm bcmd}}(t)
\right|_{t=0}
=
\left\langle
v_{\rm c}^{},
\mathcal K_{\rm bcmd}^{} v_{\rm c}^{}
\right\rangle_{W}
=
\frac{1}{m^{}} \left\langle v_{\rm c}^{}, F_{\rm c}^{} \right\rangle_{W}
=
0 ,
\label{eq:Cvv_first_derivative_zero_rewrite}
\end{equation}
again by parity.
Furthermore, since \(v_{\rm c}^{}\) is odd under \(\Theta\) whereas
\(\mathcal K_{\rm bcmd}^{} v_{\rm c}^{} = F_{\rm c}^{}/m\) is even,
Eq.~\eqref{eq:BCMD_reversibility_identity} gives
\begin{align}
\left.
\frac{d^{2}}{dt^{2}}
C_{vv,P}^{\,{\rm bcmd}}(t)
\right|_{t=0}
&=
\left\langle
v_{\rm c}^{},
\mathcal K_{\rm bcmd}^{2} v_{\rm c}^{}
\right\rangle_{W}
\nonumber\\
&=
-
\left\langle
\mathcal K_{\rm bcmd}^{} v_{\rm c}^{},
\mathcal K_{\rm bcmd}^{} v_{\rm c}^{}
\right\rangle_{W}
\nonumber\\
&=
-
\frac{1}{m^{2}}
\left\langle
F_{\rm c}^{2}
\right\rangle_{W} .
\label{eq:Cvv_second_derivative_force_rewrite}
\end{align}
Hence
\begin{equation}
C_{vv,P}^{\,{\rm bcmd}}(t)
=
\frac{1}{\beta^{}m^{}}
-
\frac{t^{2}}{2m^{2}}
\left\langle
F_{\rm c}^{2}
\right\rangle_{W}
+
O(t^{3}) .
\label{eq:Cvv_short_time_force_rewrite}
\end{equation}
Using Eq.~\eqref{eq:Cvv_from_Cxx_rewrite}, this implies
\begin{equation}
C_{xx,P}^{\,{\rm bcmd}}(t)
=
C_{xx,P}^{\,{\rm bcmd}}(0)
-
\frac{t^{2}}{2\beta^{}m^{}}
+
\frac{t^{4}}{24m^{2}}
\left\langle
F_{\rm c}^{2}
\right\rangle_{W}
+
O(t^{5}) .
\label{eq:Cxx_short_time_force_rewrite}
\end{equation}

Now \(F_{\rm c}^{}(\mathbf{q})\) is itself a position-diagonal path estimator,
Eq.~\eqref{eq:Fc_bead_average_force_rewrite}.
Therefore its equal-time second moment is exact in the
fine-discretization limit:
\begin{equation}
\lim_{P\to\infty}
\left\langle
F_{\rm c}^{2}
\right\rangle_{W}
=
C_{FF}^{\,{\rm K}}(0) ,
\quad
\hat F^{} \equiv -V'(\hat x^{}) .
\label{eq:Fc2_exact_limit_rewrite}
\end{equation}
Equivalently,
\begin{equation}
\left\langle
F_{\rm c}^{2}
\right\rangle_{W}
=
\frac{1}{Z_P^{}}
\int d\mathbf{q}\,
\bar\rho_{\beta,P}^{\,{\rm nm}}(\mathbf{q})
\left[
\frac{1}{P}
\sum_{s=1}^{P}
V'\!\left( x_s^{}(\mathbf{q}) \right)
\right]^{2},
\label{eq:force_square_path_average_rewrite}
\end{equation}
which tends to the exact Kubo equal-time force-force correlation as
\(P\to\infty\).

Thus the BCMD short-time expansions become
\begin{equation}
C_{vv}^{\,{\rm bcmd}}(t)
=
\frac{1}{\beta^{}m^{}}
-
\frac{t^{2}}{2m^{2}}
C_{FF}^{\,{\rm K}}(0)
+
O(t^{3}) ,
\label{eq:Cvv_short_time_exact_limit_rewrite}
\end{equation}
and
\begin{equation}
C_{xx}^{\,{\rm bcmd}}(t)
=
C_{xx}^{\,{\rm K}}(0)
-
\frac{t^{2}}{2\beta^{}m^{}}
+
\frac{t^{4}}{24m^{2}}
C_{FF}^{\,{\rm K}}(0)
+
O(t^{5}) .
\label{eq:Cxx_short_time_exact_limit_rewrite}
\end{equation}
Therefore BCMD reproduces the exact Kubo short-time coefficients through
\(O(t^{4})\) for \(C_{xx}\) and through \(O(t^{2})\) for the induced
\(C_{vv}\).  The first possible BCMD-specific corrections enter at
\(O(t^{5})\) and \(O(t^{3})\), respectively.

\subsection{Harmonic oscillator}

For the one-dimensional harmonic oscillator,
\begin{equation}
V(x^{})
=
\frac{1}{2} m^{} \omega^{2} x^{2} ,
\label{eq:harmonic_potential_rewrite}
\end{equation}
the ring-polymer normal modes decouple, and the effective potential takes
the form
\begin{equation}
V_{\rm eff}^{}(\mathbf{q})
=
\frac{1}{2} m^{} \omega^{2} q_{\rm c}^{2}
+
\sum_{\alpha=2}^{P}
\frac{1}{2} m^{}
\left(
\omega^{2} + \Omega_{\alpha}^{2}
\right)
q_{\alpha}^{2} ,
\label{eq:Veff_harmonic_rewrite}
\end{equation}
where
\begin{equation}
\Omega_{\alpha}^{2}
\equiv
\frac{k_{\alpha}^{}}{m^{}}
=
\omega_P^{2}\lambda_{\alpha}^{}
\label{eq:Omega_harmonic_definition}
\end{equation}
is the ring-polymer spring frequency squared of mode \(\alpha\) in the
original BCMD normal-mode convention.
Accordingly, the physical-potential part is
\begin{equation}
\phi(\mathbf{q})
=
\frac{1}{2} m^{} \omega^{2} q_{\rm c}^{2}
+
\frac{1}{2} m^{} \omega^{2}
\sum_{\alpha=2}^{P} q_{\alpha}^{2} .
\label{eq:phi_harmonic_full_rewrite}
\end{equation}
In particular, the centroid force is
\begin{equation}
-
\frac{\partial \phi}{\partial q_{\rm c}^{}} =
- m^{} \omega^{2} q_{\rm c}^{} ,
\label{eq:phi_harmonic_centroid_force_rewrite}
\end{equation}
so the centroid equations of motion are exactly the physical harmonic
equations,
\begin{equation}
\dot q_{\rm c}^{}
=
\frac{p_{\rm c}^{}}{m^{}} ,
\quad
\dot p_{\rm c}^{}
=
-
m^{} \omega^{2} q_{\rm c}^{} .
\label{eq:centroid_harmonic_eqns_rewrite}
\end{equation}
The non-centroid overdamped modes are dynamically decoupled from the
centroid and therefore do not affect \(C_{xx}\) or the induced
\(C_{vv}\).

Consequently,
\begin{equation}
C_{xx,P}^{\,{\rm bcmd}}(t)
=
\left\langle
q_{\rm c}^{}(0) q_{\rm c}^{}(t)
\right\rangle
=
\frac{1}{\beta^{} m^{} \omega^{2}}
\cos(\omega t) ,
\label{eq:Cxx_harmonic_exact_rewrite}
\end{equation}
and
\begin{equation}
C_{vv,P}^{\,{\rm bcmd}}(t)
=
\left\langle
v_{\rm c}^{}(0) v_{\rm c}^{}(t)
\right\rangle
=
\frac{1}{\beta^{} m^{}}
\cos(\omega t) .
\label{eq:Cvv_harmonic_exact_rewrite}
\end{equation}
These expressions are independent of \(P\), and therefore in the
fine-discretization limit they coincide with the exact harmonic
Kubo-transformed results.
Thus BCMD is exact for \(C_{xx}\) and the induced \(C_{vv}\) in the
harmonic case, just as in the standard harmonic analysis of CMD and
RPMD.

\subsection{Relation to PACMD and CMD}

The present formulation also makes the relation between BCMD, PACMD, and CMD
transparent.  PACMD remains a centroid-dynamics formulation in which the
non-centroid modes are auxiliary sampling variables: their fictitious masses
are chosen to realize an adiabatic separation, and thermostatting is used for
the fast internal modes.\cite{HoneRosskyVoth2006,
HabershonFanourgakisManolopoulos2008,PerezTuckermanMuser2009}
BCMD instead retains the path coordinates explicitly and propagates the
non-centroid coordinates by first-order Brownian dynamics, with no inertial
non-centroid normal-mode frequencies.  Accordingly, for general position-diagonal observables,
BCMD uses the beadwise path estimator
\begin{equation}
\bar A_P^{\,{\rm nm}}(\mathbf q)
=
\frac{1}{P}\sum_{s=1}^{P}A\!\left(x_s(\mathbf q)\right),
\label{eq:bcmd_beadwise_estimator_relation}
\end{equation}
rather than reducing the observable to the centroid alone.  As shown above,
this construction gives the exact Kubo equal-time correlation in the
\(P\to\infty\) limit and therefore the corresponding exact zeroth spectral
moment.  For a linear position observable the beadwise estimator reduces to
the centroid, whereas for a nonlinear observable the distinction is
essential.

To describe the CMD limit, we define
\begin{equation}
\mathbf{q}_{\rm nc} \equiv \left( q_{2}^{}, \dots, q_{P}^{} \right)^{\mathsf T} .
\label{eq:qnc_def}
\end{equation}
The centroid marginal is
\begin{equation}
\bar\rho_{\beta,P,{\rm c}}^{}(q_{\rm c}^{})
=
\int d\mathbf{q}_{\rm nc}\, \bar\rho_{\beta,P}^{\,{\rm nm}}(q_{\rm c}^{},\mathbf{q}_{\rm nc}) ,
\label{eq:rho_c_marginal}
\end{equation}
and the corresponding centroid free energy is
\begin{equation}
F_{{\rm c},P}^{}(q_{\rm c}^{})
=
- \frac{1}{\beta^{}} \ln \bar\rho_{\beta,P,{\rm c}}^{}(q_{\rm c}^{})
+ {\rm const.}
\label{eq:Fcp}
\end{equation}
Its derivative is
\begin{equation}
\frac{dF_{{\rm c},P}^{}}{dq_{\rm c}^{}} =
\frac{
\displaystyle \int d\mathbf{q}_{\rm nc}\, \bar\rho_{\beta,P}^{\,{\rm nm}}(q_{\rm c}^{},\mathbf{q}_{\rm nc}) \,
\dfrac{\partial \phi}{\partial q_{\rm c}^{}}
}{
\displaystyle \int d\mathbf{q}_{\rm nc}\, \bar\rho_{\beta,P}^{\,{\rm nm}}(q_{\rm c}^{},\mathbf{q}_{\rm nc})
} .
\label{eq:centroid_mean_force_identity}
\end{equation}
Thus the gradient of \(F_{{\rm c},P}^{}\) is the conditional equilibrium average of
\(\partial\phi/\partial q_{\rm c}\), or equivalently the negative of the
conditional mean centroid force.  In the formal adiabatic limit in which the
non-centroid modes equilibrate infinitely rapidly at fixed \(q_{\rm c}^{}\),
the slow centroid dynamics therefore reduces to
\begin{align}
\dot q_{\rm c}^{} &= \frac{p_{\rm c}^{}}{m^{}},
\label{eq:cmd_limit_eqn_a} \\
\dot p_{\rm c}^{} &= - \frac{dF_{{\rm c},P}^{}}{dq_{\rm c}^{}} .
\label{eq:cmd_limit_eqn_b}
\end{align}
This is precisely the CMD mean-force dynamics at finite \(P\).  Finally,
\(F_{{\rm c},P}^{}\to F_{\rm c}^{}\) as \(P\to\infty\).  Thus CMD is recovered,
at the level of the slow centroid dynamics, by adiabatically eliminating the
explicit path-resolved non-centroid dynamics, whereas BCMD retains the
non-centroid dynamics explicitly as a defined stochastic part of the
dynamical approximation.

\section{Numerical algorithm for BCMD}
\label{sec:numalgo_bcmd}

\subsection{Velocity-refresh representation of Brownian dynamics}
\label{subsec:numalgo_velocity_refresh}

Consider the overdamped Langevin equation
\begin{equation}
\Gamma \dot q
=
F(q)
+
\sqrt{\frac{2\Gamma}{\beta}}\,\xi(t).
\label{eq:numalgo_generic_brownian}
\end{equation}
Its Euler--Maruyama update over a time interval \(\Delta t\) is
\begin{equation}
\Delta q
=
\frac{\Delta t}{\Gamma}F(q)
+
\sqrt{\frac{2\Delta t}{\beta\Gamma}}\,R,
\quad
R\sim\mathcal N(0,1).
\label{eq:numalgo_em}
\end{equation}

The same increment is obtained from a velocity-Verlet step if an auxiliary
velocity is completely randomized before every interval,
\begin{equation}
v
\sim
\mathcal N\!\left(0,\frac{1}{\beta\mu}\right),
\end{equation}
and the auxiliary mass is chosen as
\begin{equation}
\mu=\frac{\Gamma\Delta t}{2}
.
\label{eq:numalgo_mass_mapping}
\end{equation}
Indeed, the first half kick followed by the coordinate drift gives
\begin{equation}
\Delta q
=
v\Delta t
+
\frac{\Delta t^2}{2\mu}F(q),
\end{equation}
whose first two moments are identical to those of
Eq.~\eqref{eq:numalgo_em}.  This is a fully randomized Andersen-type velocity-refresh representation of the Euler--Maruyama discretization of Brownian dynamics.\cite{Andersen1980}
It is also analogous to a hybrid Monte Carlo (HMC) step without the Metropolis
accept/reject procedure,\cite{DuaneKennedyPendletonRoweth1987} although here
the auxiliary velocity is introduced to generate the Brownian dynamics rather
than to sample an equilibrium distribution.

For BCMD,
\begin{equation}
\Gamma_\alpha=m\gamma_\alpha,
\end{equation}
and hence
\begin{equation}
\mu_\alpha
=
\frac{m\gamma_\alpha\Delta t}{2},
\quad
\alpha=2,\ldots,P.
\label{eq:numalgo_mu}
\end{equation}
The important point is that \(\mu_\alpha\) is determined by the velocity
refresh interval \(\Delta t\).  We use the corresponding auxiliary
momenta
\begin{equation}
p_\alpha \equiv \mu_\alpha v_\alpha,
\quad
\alpha=2,\ldots,P.
\label{eq:numalgo_aux_momenta}
\end{equation}

\subsection{MTS-BCMD algorithm}
\label{subsec:numalgo_mts}

Between successive velocity refreshes, we propagate the centroid and the
auxiliary non-centroid variables with the pseudo-Hamiltonian
\begin{equation}
H_{\rm ps}
=
\frac{p_{\rm c}^2}{2m}
+
\sum_{\alpha=2}^{P}
\frac{p_\alpha^2}{2\mu_\alpha}
+
V_{\rm eff}(\mathbf q).
\label{eq:numalgo_hps}
\end{equation}
We separate the free-ring-polymer part from the physical potential,
\begin{equation}
V_{\rm eff}(\mathbf q)
=
\sum_{\alpha=2}^{P}
\frac12 k_\alpha q_\alpha^2
+
\phi(\mathbf q),
\end{equation}
and write
\begin{equation}
H_{\rm ps}=H_0+\phi(\mathbf q),
\end{equation}
where
\begin{equation}
H_0
=
\frac{p_{\rm c}^2}{2m}
+
\sum_{\alpha=2}^{P}
\left[
\frac{p_\alpha^2}{2\mu_\alpha}
+
\frac12 k_\alpha q_\alpha^2
\right].
\label{eq:numalgo_h0}
\end{equation}

For any Hamiltonian \(H\), let \(\mathcal K_H\) denote the associated
Hamiltonian Liouville generator acting on phase-space functions.  We write
\(\mathcal K_{\rm ps}\), \(\mathcal K_0\), and \(\mathcal K_\phi\)
for the generators associated with \(H_{\rm ps}\), \(H_0\), and
\(\phi(\mathbf q)\), respectively.

The interval between velocity refreshes is divided into \(L\) deterministic
substeps,
\begin{equation}
\Delta t=L\delta t
,
\quad
L=1,2,\ldots .
\label{eq:numalgo_step_relation}
\end{equation}
Each substep is propagated by the symmetric RESPA splitting
\begin{equation}
e^{\delta t \mathcal K_{\rm ps}}
=
e^{\frac{\delta t}{2}\mathcal K_\phi}
e^{\delta t \mathcal K_0}
e^{\frac{\delta t}{2}\mathcal K_\phi}
+
O(\delta t^3).
\label{eq:numalgo_respa}
\end{equation}
The \(\mathcal K_\phi\) propagator gives the force kick
\begin{align}
p_{\rm c}
&\leftarrow
p_{\rm c}
-
\frac{\delta t}{2}
\frac{\partial\phi}{\partial q_{\rm c}},
\\
p_\alpha
&\leftarrow
p_\alpha
-
\frac{\delta t}{2}
\frac{\partial\phi}{\partial q_\alpha},
\end{align}
whereas \(\mathcal K_0\) can be propagated analytically:
\begin{align}
q_{\rm c}'
&=
q_{\rm c}
+
\frac{\delta t}{m}p_{\rm c},
\nonumber \\
p_{\rm c}'&=p_{\rm c},
\nonumber \\
q_\alpha'
&=
q_\alpha\cos(\omega_\alpha^{\rm ps}\delta t)
+
\frac{p_\alpha}{\mu_\alpha\omega_\alpha^{\rm ps}}
\sin(\omega_\alpha^{\rm ps}\delta t),
\nonumber \\
p_\alpha'
&=
p_\alpha\cos(\omega_\alpha^{\rm ps}\delta t)
-
\mu_\alpha\omega_\alpha^{\rm ps} q_\alpha
\sin(\omega_\alpha^{\rm ps}\delta t),
\end{align}
with
\begin{equation}
\omega_\alpha^{\rm ps}
=
\sqrt{\frac{k_\alpha}{\mu_\alpha}}.
\end{equation}
The quantity \(\omega_\alpha^{\rm ps}\) is an auxiliary frequency of
the pseudo-Hamiltonian propagation; it is not an inertial normal-mode
frequency of the BCMD non-centroid dynamics.

One MTS-BCMD cycle is therefore:

\begin{enumerate}
\item
Refresh the non-centroid velocities,
\begin{equation}
v_\alpha
\sim
\mathcal N\!\left(
0,\frac{1}{\beta\mu_\alpha}
\right),
\quad
\alpha=2,\ldots,P,
\end{equation}
and set \(p_\alpha=\mu_\alpha v_\alpha\).

\item
Propagate Eq.~\eqref{eq:numalgo_respa} \(L\) times with time step
\(\delta t\).

\item
Repeat the velocity refresh.
\end{enumerate}

The centroid momentum is never randomized and is propagated continuously by
its physical Newton equation.  The choice \(L=1\) reduces to the original
BCMD algorithm,\cite{Shiga2022BCMD} while \(L>1\) resolves the deterministic
pseudo-Hamiltonian trajectory more accurately without increasing the
frequency of the Brownian velocity refresh.

A related reduction in the frequency of stochastic updates was noted by Ceriotti \textit{et al.} in their discussion of stochastic thermostatting for path-integral molecular dynamics, where the thermostat propagator may be applied less frequently than the deterministic propagation.\cite{CeriottiParrinelloMarklandManolopoulos2010} In the path-integral Langevin equation (PILE) formulation, however, the ring-polymer masses are independent of the thermostat interval. The situation is different in the present MTS-BCMD algorithm. Equation~\eqref{eq:numalgo_mu} gives \(\mu_\alpha=m\gamma_\alpha\Delta t/2\), and, using Eq.~\eqref{eq:numalgo_step_relation}, \(\mu_\alpha=m\gamma_\alpha L\delta t/2\). Thus, the fictitious mass \(\mu_\alpha\) itself depends linearly on the stochastic refresh interval \(\Delta t=L\delta t\). This dependence originates from the \(\sqrt{\Delta t}\) scaling of the stochastic displacement in the discrete overdamped Langevin equation: matching this Brownian increment to the displacement generated by a completely randomized auxiliary velocity requires \(\mu_\alpha\) to scale linearly with \(\Delta t\).

\subsection{Choice of time steps}
\label{subsec:numalgo_timestep}

The two time steps control different numerical errors.  The inner step
\(\delta t\) controls the deterministic RESPA propagation.  Since
\(H_{\rm ps}\) is conserved between velocity refreshes in the continuous
auxiliary dynamics, its deterministic change provides a direct convergence
test.  We define
\begin{equation}
\mathbf p_{\rm nc}
\equiv
(p_2,\ldots,p_P)^{\mathsf T},
\qquad
\mathbf z
\equiv
(p_{\rm c},\mathbf p_{\rm nc},\mathbf q).
\label{eq:numalgo_z_state}
\end{equation}
If \(\mathbf z_n^+\) is the state immediately after a velocity refresh
and \(\mathbf z_{n+1}^-\) the state immediately before the next refresh, we
define
\begin{equation}
\Delta H_{{\rm ps},n}^{\rm det}
=
H_{\rm ps}(\mathbf z_{n+1}^-)
-
H_{\rm ps}(\mathbf z_n^+).
\label{eq:numalgo_energy_det}
\end{equation}
The stochastic energy change caused by the velocity refresh is excluded.
A practical criterion is
\begin{equation}
{\rm rms}
\left(
\Delta H_{\rm ps}^{\rm det}
\right)
\ll
\beta^{-1},
\label{eq:numalgo_energy_beta}
\end{equation}
together with the absence of systematic accumulated drift.

By contrast, the outer interval \(\Delta t\) controls the discretization of
the non-centroid Brownian dynamics.  In BCMD,
\begin{equation}
\tau_\alpha
=
\frac{m\gamma_\alpha}{k_\alpha}
=
\beta\hbar,
\quad
\alpha=2,\ldots,P,
\end{equation}
so that the refresh interval should satisfy
\begin{equation}
\Delta t\ll\beta\hbar
.
\end{equation}
Thus \(\delta t\) is chosen from pseudo-Hamiltonian conservation, whereas
\(\Delta t\) is chosen from convergence of the Brownian dynamics.



\section{Numerical tests}
\label{sec:numerical_tests}

The numerical tests address three separate issues.  Figure~\ref{fig:pathqdo_validation}
tests the finite-\(P\) path-QDO representation itself.
Figure~\ref{fig:mts_1d}, Fig.~\ref{fig:mts_water}, and
Table~\ref{tab:mts_energy} test the MTS implementation introduced in
Sec.~\ref{subsec:numalgo_mts}, in particular the separation of the Brownian
velocity-refresh interval \(\Delta t\) from the deterministic inner step
\(\delta t\).  Figures~\ref{fig:bcmd_water_comparison} and
\ref{fig:bcmd_water_comparison_2} then examine the mode-dependent spectral
behavior of BCMD for q-TIP4P/F ice and water.

\subsection{Path-QDO validation}
\label{subsec:pathqdo_validation}

We first isolate the imaginary-time discretization entering the path-QDO.
Specifically, we evaluate the finite-\(P\) symmetric-Trotter approximation to the source-derivative expression in Eq.~\eqref{eq:Kubo_from_source_derivative}, retaining the finite-\(P\) endpoint terms appearing in Eq.~\eqref{eq:open_source_full_derivative}.
Its convergence with increasing \(P\) therefore tests the imaginary-time discretization underlying the path-QDO representation and its continuum-limit form in Eq.~\eqref{eq:Kubo_path_QDO_final_rp}.
The source derivative is evaluated analytically at finite $P$, rather than by a finite difference in the source strength.

The
real-time propagation is evaluated quantum mechanically, so that no BCMD
dynamical approximation is involved.  We use the three one-dimensional models,\cite{CraigManolopoulos2004,Shiga2022BCMD}
\begin{align}
 V_{\rm m}(x)&=\frac12x^2+\frac1{10}x^3+\frac1{10}x^4,
 & A(x)&=x,\nonumber\\
 V_{\rm q}(x)&=\frac12x^4,
 & A(x)&=x,\nonumber\\
 V_{\rm h}(x)&=\frac12x^2,
 & A(x)&=x^2,
\end{align}
with \(m=\hbar=1\) at \(\beta=1\) and 8.  The last model also tests the
nonlinear observable \(A(x)=x^2\), for which the path-QDO closure does not rely
on a linear-operator restriction.

\begin{figure}[!htbp]
 \centering
 \includegraphics[width=0.98\columnwidth]{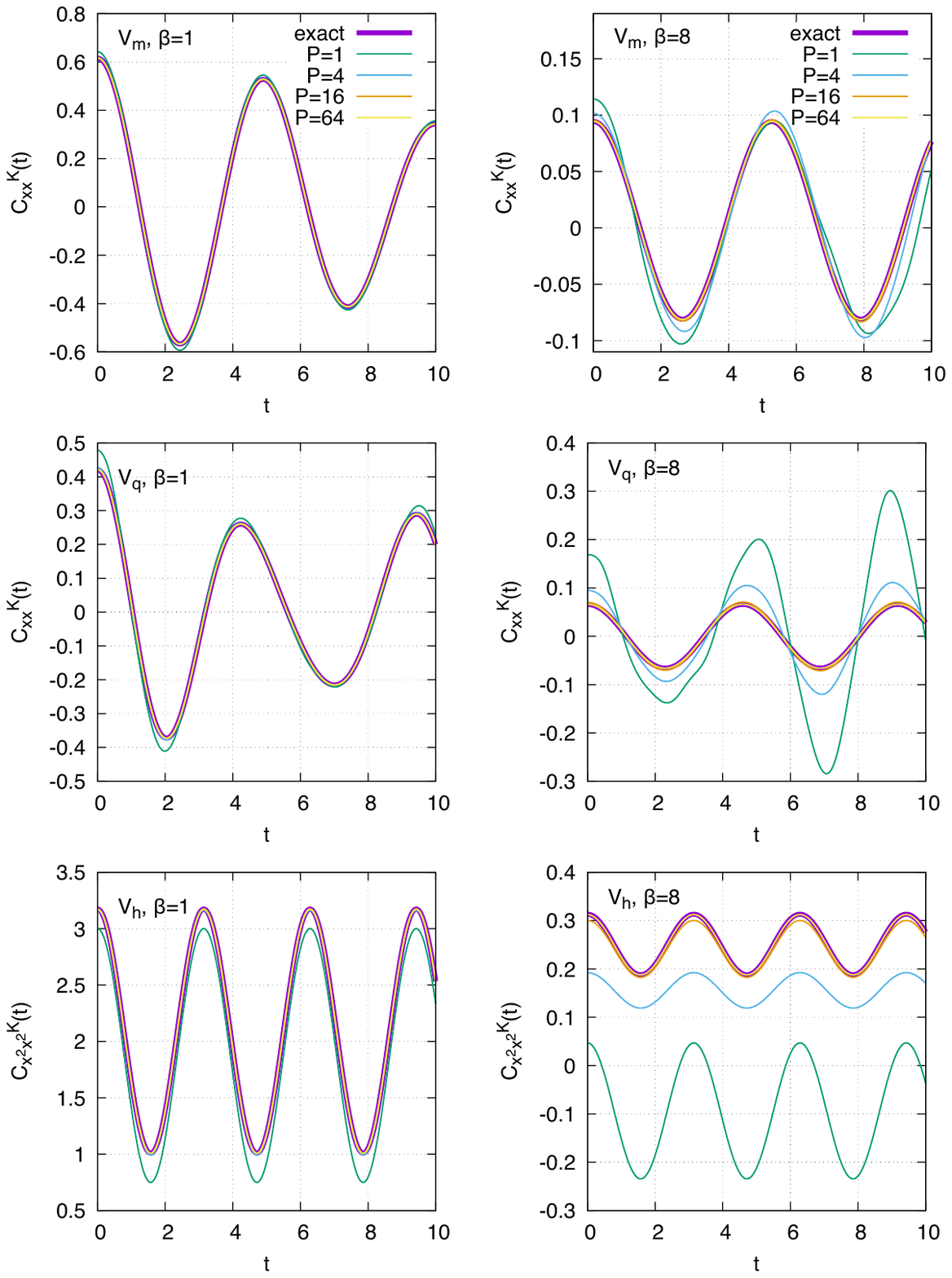}
 \caption{Finite-\(P\) validation of the path-QDO representation for the three
 one-dimensional models.  The left and right columns show \(\beta=1\) and 8,
 respectively.  The real-time propagation is exact; the \(P\) dependence
 therefore measures only the imaginary-time discretization error.}
 \label{fig:pathqdo_validation}
\end{figure}

Figure~\ref{fig:pathqdo_validation} compares the finite-\(P\)
Kubo-transformed autocorrelation functions with exact eigenstate results.  In
all six model--temperature combinations the finite-\(P\) curves converge
smoothly to the exact result.  For the bead number \(P=4\beta\) used in the
original BCMD calculations, the normalized RMS deviation over the plotted
time interval is below \(2.1\%\).  The finite-bead representation used in the
dynamical tests below is therefore already well converged.

The results presented here validate the path-QDO formulation and
its convergence to the exact path-resolved representation. They should be
distinguished from the validation of the BCMD dynamical approximation, which
is examined separately below.

\subsection{Validation of the MTS-BCMD integrator}
\label{subsec:mts_validation}

\begin{figure}[!htbp]
 \centering
 \includegraphics[width=0.98\columnwidth]{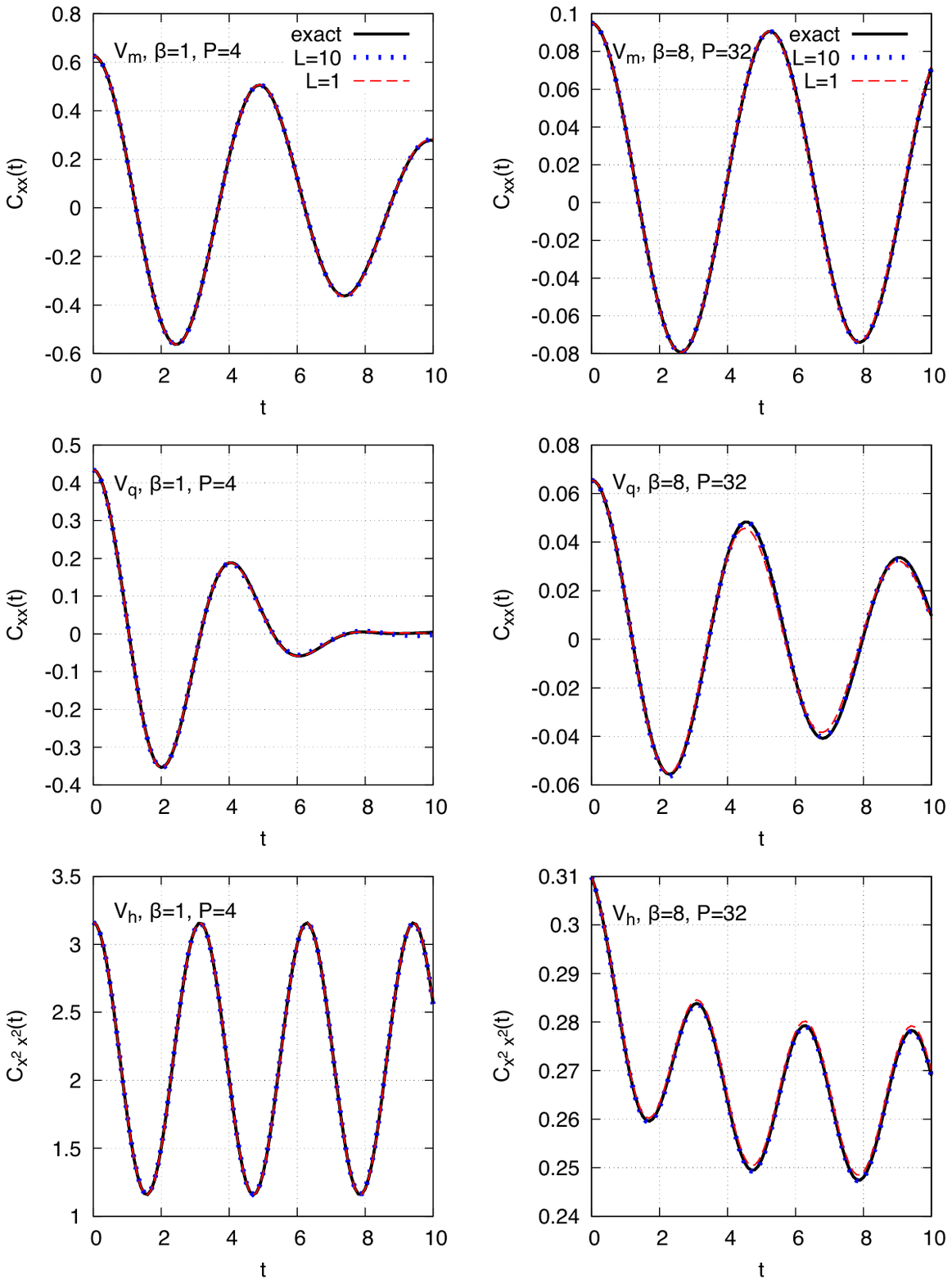}
 \caption{MTS-BCMD test for the same one-dimensional models as in
 Fig.~\ref{fig:pathqdo_validation}.  The left and right columns show
 \(\beta=1\) (\(P=4\)) and \(\beta=8\) (\(P=32\)), respectively.  The
 \(L=1\) and \(L=10\) calculations use the same Brownian refresh interval
 \(\Delta t=0.10\), corresponding to deterministic inner steps
 \(\delta t=0.10\) and \(0.01\).  Exact quantum correlation functions are
 shown as references.}
 \label{fig:mts_1d}
\end{figure}

The purpose of the MTS test is to separate deterministic integration error
from the discretization of the Brownian dynamics.  We therefore keep the
velocity-refresh interval fixed at \(\Delta t=0.10\) and change only the
number of deterministic substeps.  As shown in Fig.~\ref{fig:mts_1d}, the
\(L=1\) and \(L=10\) BCMD correlations are nearly indistinguishable.  Thus,
once \(\Delta t\) is fixed, reducing \(\delta t=\Delta t/L\) improves the
deterministic propagation without changing the intended BCMD dynamics.

A more stringent numerical test is provided by condensed-phase q-TIP4P/F
water.\cite{HabershonMarklandManolopoulos2009}  Hexagonal ice Ih at 150 K was
represented by 96 water molecules in an orthorhombic cell of dimensions
\(13.5575\times15.6530\times14.7588\)~\AA.  The oxygen atoms were placed on
the ice-Ih lattice, and a single proton-disordered configuration was
constructed by assigning the hydrogen atoms subject to the Bernal--Fowler ice
rules.  The resulting proton-disordered configuration had a negligible net cell dipole.  Liquid water
at both 300 and 600 K was represented by 128 water molecules in a cubic cell
of side length 15.6359~\AA.  Three-dimensional periodic boundary conditions
were used throughout.  The Lennard--Jones interaction was smoothly switched
off between 8.144 and 9.644~\AA, and the Coulomb interaction was evaluated by
Ewald summation with a target numerical accuracy of \(10^{-8}\) hartree.

The classical and path-integral ensembles used to initialize MD and BCMD were
equilibrated independently at each temperature.  For classical MD, 100 configurations were sampled from an
equilibrated canonical-ensemble (NVT) trajectory and used to initiate 100 separate 5 ps
microcanonical (NVE) production trajectories.  For BCMD, 100 path
configurations were sampled from an equilibrated NVT PIMD trajectory and used
to initiate 100 separate 5 ps BCMD production trajectories.  No additional
thermostat was applied during the BCMD production trajectories beyond the
intrinsic stochastic refresh of the non-centroid variables.  The number of
beads was \(P=64\), 32, and 16 for ice at 150 K, liquid water at 300 K, and
liquid water at 600 K, respectively.

Table~\ref{tab:mts_energy} reports deterministic energy-conservation
diagnostics for additional fixed-\(\Delta t\) MTS tests.  Consistently with
Sec.~\ref{subsec:numalgo_timestep}, stochastic energy changes produced by the
non-centroid velocity refresh are excluded.  At fixed \(\Delta t\), increasing
\(L\) reduces the deterministic drift by approximately one to two orders of
magnitude and strongly reduces the detrended root-mean-square (RMS)
fluctuation.  These tests demonstrate that the inner step can be refined
independently of the Brownian refresh interval.

\begin{table}[!htbp]
\centering
\scriptsize
\setlength{\tabcolsep}{2.2pt}
\caption{Energy-conservation statistics for the q-TIP4P/F MTS tests.  The
drift is in meV/(\(\mu\)s atom) and the detrended RMS fluctuation is in
meV/atom.  For BCMD, \(\Delta t\) is the velocity-refresh interval and \(L\)
is the number of deterministic inner steps, so that
\(\delta t=\Delta t/L\).  The stochastic energy change associated with each
velocity refresh is excluded from the BCMD statistics.  For MD, the
tabulated \(\Delta t\) is the ordinary integration time step and the
statistics refer to the ordinary Hamiltonian energy.}
\label{tab:mts_energy}
\begin{tabular}{llccccrr}
\hline\hline
method & system & \(T\) [K] & \(P\) & \(\Delta t\) [fs] & \(L\) & drift & RMS \\
\hline
BCMD & ice       & 150 & 64 & 0.25  &  1 & 2095 & 365 \\
BCMD & ice       & 150 & 64 & 0.25  & 10 &   22 &  49 \\
BCMD & ice       & 150 & 64 & 0.125 &  1 &  530 & 232 \\
BCMD & ice       & 150 & 64 & 0.125 & 10 &    8 &  26 \\
MD   & ice       & 150 &  1 & 0.25  & --- &    0 &   7 \\
MD   & ice       & 150 &  1 & 0.125 & --- &    0 &   2 \\
\hline
BCMD & water     & 300 & 32 & 0.25  &  1 & 1697 & 475 \\
BCMD & water     & 300 & 32 & 0.25  & 10 &   17 &  36 \\
BCMD & water     & 300 & 32 & 0.125 &  1 & 1030 & 344 \\
BCMD & water     & 300 & 32 & 0.125 & 10 &    4 &  33 \\
MD   & water     & 300 &  1 & 0.25  & --- &    0 &  12 \\
MD   & water     & 300 &  1 & 0.125 & --- &    0 &   2 \\
\hline
BCMD & hot water & 600 & 16 & 0.25  &  1 & 1232 & 577 \\
BCMD & hot water & 600 & 16 & 0.25  &  5 &  129 & 180 \\
BCMD & hot water & 600 & 16 & 0.125 &  1 &  264 & 267 \\
BCMD & hot water & 600 & 16 & 0.125 & 10 &    8 &  44 \\
MD   & hot water & 600 &  1 & 0.25  & --- &    0 &  25 \\
MD   & hot water & 600 &  1 & 0.125 & --- &    0 &   9 \\
\hline\hline
\end{tabular}
\end{table}

For the production BCMD spectra, the deterministic inner step was fixed at
\(\delta t=0.125\) fs.  Calculations with \(L=1\) and \(L=10\) therefore
corresponded to Brownian refresh intervals of \(\Delta t=0.125\) and
1.25 fs, respectively.  The dipole estimator was the bead average of the
total q-TIP4P/F dipole,
\begin{equation}
\overline{\boldsymbol{\mu}}_P(t)
=
\frac{1}{P}
\sum_{s=1}^{P}
\boldsymbol{\mu}\!\left(\mathbf{R}_s(t)\right),
\label{eq:dipole_bead_average}
\end{equation}
where \(\mathbf R_s(t)\) denotes the full Cartesian configuration of bead
\(s\), and \(\boldsymbol\mu(\mathbf R_s)\) is the total q-TIP4P/F dipole
of the simulation cell for that bead.  The autocorrelation function was
\begin{equation}
C_{\mu\mu}(t)
=
\left\langle
\overline{\boldsymbol{\mu}}_P(0)
\cdot
\overline{\boldsymbol{\mu}}_P(t)
\right\rangle .
\label{eq:dipole_autocorrelation}
\end{equation}
Up to a frequency-independent prefactor, the infrared spectrum was evaluated
as
\begin{equation}
I(\omega)
\propto
\beta\omega^2
\int_{-\infty}^{\infty}dt\,
e^{i\omega t}C_{\mu\mu}(t).
\label{eq:ir_spectrum_definition}
\end{equation}
The correlation functions were averaged over the 100 production trajectories
before Fourier transformation.  The resulting spectra were smoothed by
convolution with a Gaussian of 10~cm\(^{-1}\) full width at half maximum
(FWHM).  The classical MD reference spectra were obtained from their separate
NVT equilibrations and 100 NVE production trajectories using the same cells,
trajectory length, and spectral processing.

\begin{figure}[!htbp]
 \centering
 \includegraphics[width=0.98\columnwidth]{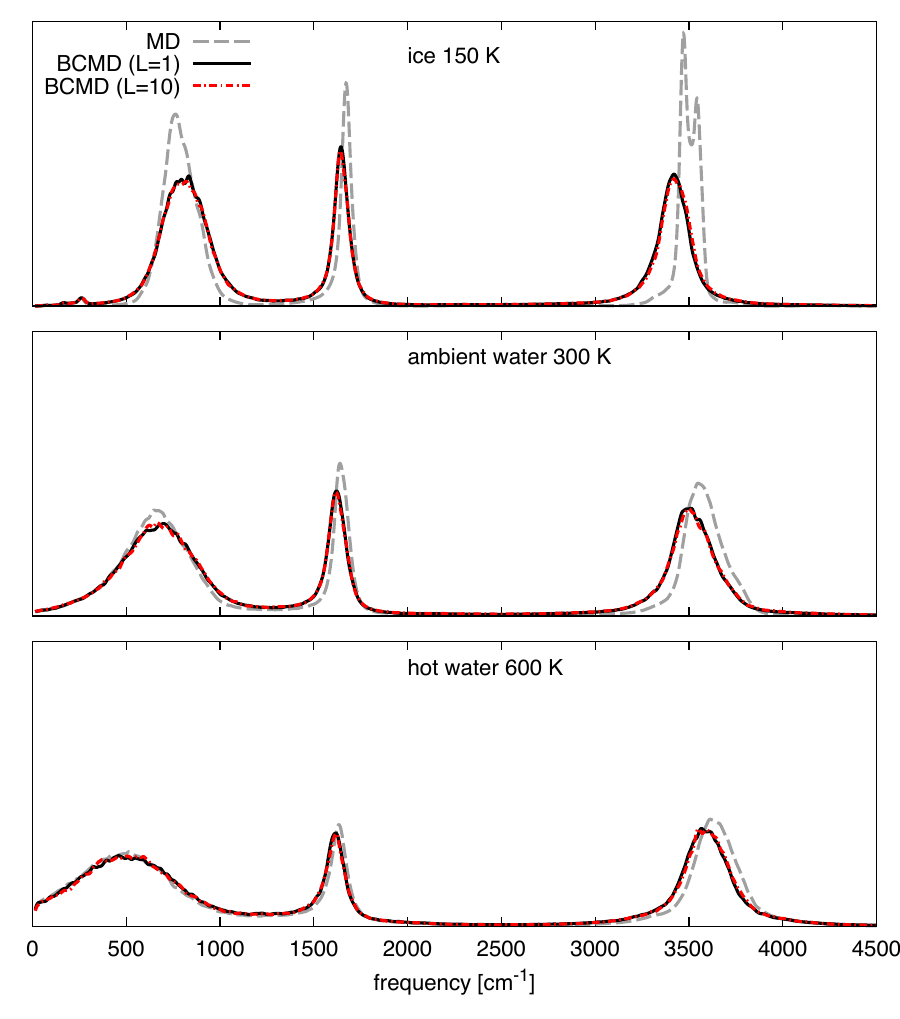}
 \caption{Infrared spectra of q-TIP4P/F ice at 150 K, ambient water at 300 K,
 and hot water at 600 K.  The two BCMD calculations use
 \(\delta t=0.125\) fs with \(L=1\) and \(L=10\), corresponding to
 \(\Delta t=0.125\) and 1.25 fs, respectively.  Classical MD spectra obtained
 with the same cells and spectral processing are shown for reference.}
 \label{fig:mts_water}
\end{figure}

Figure~\ref{fig:mts_water} compares the two BCMD production calculations.
Their near overlap shows that the spectra are insensitive to increasing the
refresh interval from 0.125 to 1.25 fs while retaining the same deterministic
inner step.  Together with Table~\ref{tab:mts_energy}, this illustrates the
main numerical advantage of the MTS formulation: the deterministic
integration accuracy and the Brownian refresh interval can be controlled
independently.

\subsection{BCMD spectra of q-TIP4P/F water and ice}
\label{subsec:bcmd_water_validation}

\begin{figure}[!htbp]
 \centering
 \includegraphics[width=0.98\columnwidth]{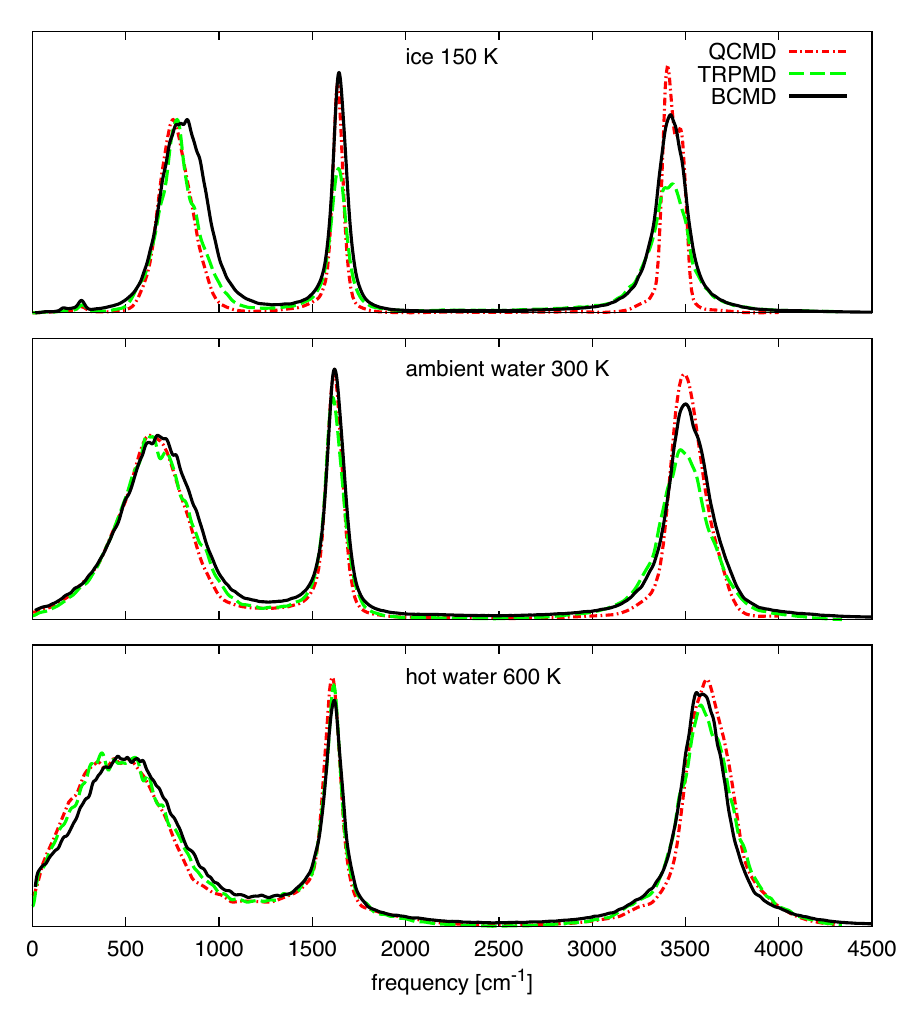}
 \caption{BCMD spectra of q-TIP4P/F ice and water compared with published QCMD
 and TRPMD spectra.  Each BCMD curve is the arithmetic average of the
 \(L=1\) and \(L=10\) spectra in Fig.~\ref{fig:mts_water}.  At 600 K, both
 the QCMD and TRPMD curves were reconstructed from Trenins, Willatt, and Althorpe.\cite{TreninsWillattAlthorpe2019}  The literature curves were
 independently rescaled to the same maximum in the 500--1000 cm\(^{-1}\)
 interval.  The comparison therefore concerns peak positions and line shapes,
 not absolute intensities.}
 \label{fig:bcmd_water_comparison}
\end{figure}

\begin{figure}[!htbp]
 \centering
 \includegraphics[width=0.98\columnwidth]{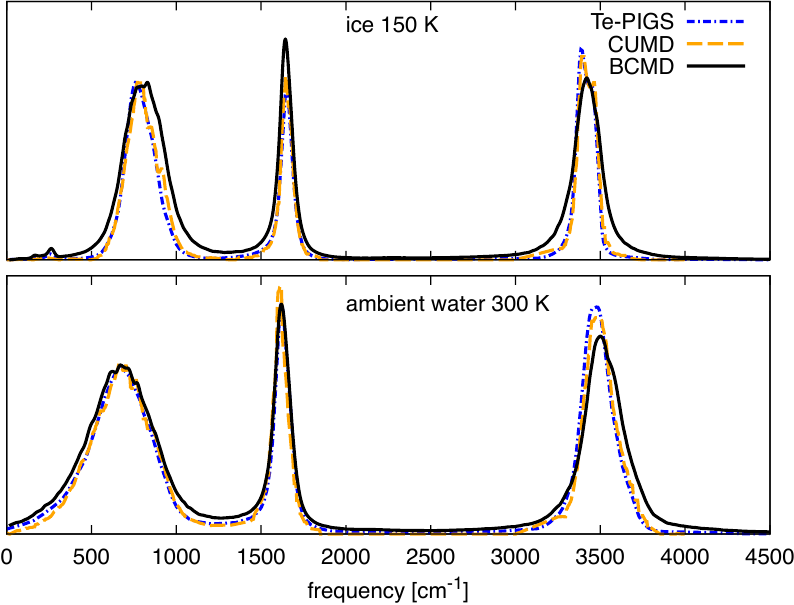}
 \caption{BCMD spectra of q-TIP4P/F ice and ambient water compared with
 published Te-PIGS and CUMD spectra.  Each BCMD curve is the arithmetic
 average of the \(L=1\) and \(L=10\) spectra in Fig.~\ref{fig:mts_water}.
 The literature curves were reconstructed from the published figures and
 independently rescaled to the same maximum in the 500--1000 cm\(^{-1}\)
 interval.}
 \label{fig:bcmd_water_comparison_2}
\end{figure}

Figures~\ref{fig:bcmd_water_comparison} and
\ref{fig:bcmd_water_comparison_2} compare BCMD spectra for ice at 150 K,
ambient water at 300 K, and hot water at 600 K with published q-TIP4P/F
spectra obtained using other approximate quantum-dynamics methods.  In the
original BCMD study,\cite{Shiga2022BCMD} the condensed-phase spectral
benchmark was centered on ambient liquid water.  Here we extend the assessment
of BCMD to low-temperature ice and high-temperature liquid water in addition
to ambient water.

For the comparisons below, the BCMD curve at each state point was taken as
the arithmetic average of the \(L=1\) and \(L=10\) spectra in
Fig.~\ref{fig:mts_water}.  The QCMD spectra at 150 and 300 K are taken from
the revised calculations of Trenins, Haggard, and
Althorpe,\cite{TreninsHaggardAlthorpe2022} whereas the 600 K QCMD spectrum is
reconstructed from Trenins, Willatt, and Althorpe.\cite{TreninsWillattAlthorpe2019}
The TRPMD spectra at 150 and 300 K are taken from the q-TIP4P/F calculations
of Rossi \textit{et al.},\cite{RossiLiuPaesaniBowmanCeriotti2014} whereas the
600 K TRPMD spectrum is likewise reconstructed from
Trenins, Willatt, and Althorpe.\cite{TreninsWillattAlthorpe2019}
The Te-PIGS spectra at 150 and 300 K are taken from Musil
\textit{et al.},\cite{MusilZaporozhetsNoeClementiKapil2022} and the CUMD
spectra are taken from Hasegawa.\cite{Hasegawa2023CUMD}  All literature
curves were reconstructed from the corresponding published figures; no QCMD,
TRPMD, Te-PIGS, or CUMD calculations were performed in the present work.
Since each literature curve was independently rescaled, the comparison is
restricted to peak positions and line shapes rather than absolute
intensities.

In the bending and OH-stretch regions, the BCMD peaks lie close to the QCMD
and TRPMD peaks in Fig.~\ref{fig:bcmd_water_comparison}.  In the librational
region, particularly for ice at 150 K, the BCMD peak lies closer to the
Te-PIGS and CUMD peaks in Fig.~\ref{fig:bcmd_water_comparison_2}.  These
statements refer only to the relative positions of the spectra assembled
here; none of the other methods is treated as an exact reference.

The librational band deserves separate attention.  Relative to the classical
MD spectra in Fig.~\ref{fig:mts_water}, BCMD shifts the librational peak
slightly to higher frequency.  The Te-PIGS and CUMD spectra displayed in
Fig.~\ref{fig:bcmd_water_comparison_2} show shifts in the same direction.
This similarity is noteworthy because the calculations are based on different
dynamical approximations.  It does not establish that the BCMD shift is
quantitatively correct, nor does it imply any statement by the authors of
Te-PIGS or CUMD about the origin of the BCMD shift.  It shows only that a
small blue shift of the librational peak relative to classical MD is not
unique to the BCMD spectrum.

The line shapes should be interpreted independently of the peak positions.
The BCMD librational bands shown here are broader than the corresponding
Te-PIGS and CUMD bands.  Such broadening may be associated with the Brownian
treatment of the non-centroid real-time dynamics in BCMD, but the present
comparison by itself does not establish its origin.  In the OH-stretch
region, the TRPMD spectra shown in Fig.~\ref{fig:bcmd_water_comparison} are
broader than the BCMD spectra, particularly at the lower temperatures.
This difference is consistent with the qualitatively different
non-centroid equations of motion in the two methods.  In TRPMD the
non-centroid ring-polymer normal modes retain inertial second-order dynamics,
with friction
\(\widetilde\gamma_k=2\lambda|\widetilde\omega_k|\) chosen mode by mode from
the corresponding free-ring-polymer frequency.  In the convention of Rossi,
Ceriotti, and Manolopoulos, \(\lambda=1\) gives critical damping, while their
recommended choice \(\lambda=1/2\) is underdamped.\cite{RossiCeriottiManolopoulos2014}
The internal ring-polymer vibrations are therefore damped rather than
eliminated and can indirectly broaden physical vibrational peaks.  BCMD is
qualitatively different: its non-centroid modes obey first-order overdamped
Brownian equations, with the mode-dependent friction parameters chosen so that
every non-centroid mode has the same relaxation time,
\(\tau_\alpha=m\gamma_\alpha/k_\alpha=\beta\hbar\).  The absence of
inertial non-centroid oscillations therefore provides a natural qualitative
explanation for the smaller OH-stretch broadening obtained with BCMD; this is
not intended as a quantitative decomposition of the linewidth.
Thus, within the comparisons made here, BCMD gives peak positions that fall
within the range spanned by the other approximate quantum-dynamics methods,
while its librational line shapes tend to be broader.

BCMD has also recently been incorporated as an official
path-integral propagator in CP2K version 2026.2.\cite{CP2K2026BCMD}
As BCMD begins to be used beyond its original methodological studies, the
systematic comparisons presented here may therefore provide a useful
benchmark for future applications of the method.


\section{Conclusion} \label{sec:conclusion_jcp}

\begingroup\color{black}
In this paper, we have introduced a path-resolved QDO representation
by extending the centroid-QDO construction to the full discretized
imaginary-time path.
\endgroup  For two position-diagonal observables,
the resulting path-QDO gives an exact representation of the ordinary Kubo
transform in the \(P\to\infty\) limit, expressed as an equilibrium integral
\begingroup\color{black}
over \((\mathbf{x},p_{\rm c})\) with the exact real-time quantum dynamics
retained in a conditional path-resolved estimator. BCMD can be viewed
 as a phenomenological, time-homogeneous Markov approximation to this exact reduced
propagation, with Newtonian centroid
dynamics and overdamped Brownian relaxation of the non-centroid path coordinates.
\endgroup

As a secondary numerical development, we have separated the stochastic
non-centroid refresh interval from the deterministic pseudo-Hamiltonian inner
step.  The resulting multiple-time-step (MTS) implementation allows the
deterministic integration error to be monitored through the
pseudo-Hamiltonian change between successive refreshes, while the Brownian
refresh interval is controlled independently.

The numerical tests support both aspects of the formulation.  The
finite-\(P\) path-QDO representation converges to exact one-dimensional Kubo
correlation functions, including a nonlinear position observable, and the MTS
implementation preserves the BCMD spectra while improving control of the
deterministic propagation.  For q-TIP4P/F water, BCMD was further assessed
for ice at 150 K, ambient liquid water at 300 K, and hot liquid water at
600 K, extending the condensed-phase tests beyond the ambient-water benchmark
of the original work.  These results provide a clearer theoretical framework and numerical
benchmark for applying BCMD to quantum vibrational dynamics in anharmonic
molecular and condensed-phase systems.


\begin{acknowledgments}
M.S. thanks the JSPS Grant-in-Aid for Scientific Research (Grant Numbers 24K01145, 25K01629, 26K01464, and 26H00488) for financial support. The calculations were conducted using the supercomputing facilities of Japan Atomic Energy Agency (HPE8600) and The University of Tokyo (Wisteria/BDEC-01).
\end{acknowledgments}


\section*{Data Availability}

The data that support the findings of this study are available within
 the article.

%


\end{document}